\documentclass[twocolumn]{aastex63}

\usepackage[normalem]{ulem}
\usepackage{subfigure}
\usepackage{xspace}
\usepackage[utf8]{inputenc}
\usepackage{multirow,amsmath}
\usepackage{float}
\usepackage{apjfonts}

\newcommand\msun{${M}_{\odot}$ }                   
\newcommand{\Msun}{\msun}                           
\newcommand{\BV}{Brunt-V\"{a}is\"{a}l\"{a}}         

\newcommand{\code}[1]{\texttt{#1}}
\newcommand{\mesa}{\code{MESA} }
\newcommand{\MESA}{\mesa}
\newcommand{\flash}{\code{FLASH} }
\newcommand{\FLASH}{\flash}

\shorttitle{Multidimensional Stellar Models of Massive Stars}
\shortauthors{Fields \& Couch}

\begin{document}
\title{\Large On The Development of Multidimensional Progenitor Models For Core-collapse Supernovae}

\correspondingauthor{C.~E.~Fields}
\email{fieldsc9@msu.edu}

\author[0000-0002-8925-057X]{C.~E.~Fields}
\altaffiliation{NSF Graduate Research Fellow}
\altaffiliation{Ford Foundation Predoctoral Fellow} 
\affiliation{Department of Physics and Astronomy, Michigan State University, East Lansing, MI 48824, USA}
\affiliation{Center for Theoretical Astrophysics, Los Alamos National Lab, Los Alamos, NM 87545, USA}
\affiliation{Joint Institute for Nuclear Astrophysics - Center for the Evolution of the Elements, USA}
\affiliation{Department of Computational Mathematics, Science, and Engineering, Michigan State University, East Lansing, MI 48824, USA}

\author[0000-0002-5080-5996]{Sean M.~Couch}
\affiliation{Department of Physics and Astronomy, Michigan State University, East Lansing, MI 48824, USA}
\affiliation{Department of Computational Mathematics, Science, and Engineering, Michigan State University, East Lansing, MI 48824, USA}
\affiliation{National Superconducting Cyclotron Laboratory, Michigan State University, East Lansing, MI 48824, USA}
\affiliation{Joint Institute for Nuclear Astrophysics - Center for the Evolution of the Elements, USA}

\begin{abstract}
Multidimensional hydrodynamic simulations of shell convection in massive stars 
suggest the development of aspherical perturbations that may be amplified during iron core-collapse. 
These perturbations have a crucial and qualitative impact on the delayed neutrino-driven core-collapse
supernova explosion mechanism by increasing the total stress behind the stalled shock.
In this paper, we investigate the properties of a 15 \msun model evolved in 1-,2-, and 
3-dimensions (3D) for the final $\sim$424 seconds before gravitational
instability and iron core-collapse using \texttt{MESA} and the \texttt{FLASH} simulation framework.
We find that just before collapse, our initially perturbed fully 3D model reaches angle-averaged
convective velocity magnitudes of $\approx$ 240-260 km s$^{-1}$ in the Si- and O-shell regions with 
a Mach number $\approx$ 0.06. We find the bulk of the power in the O-shell resides at large scales, 
characterized by spherical harmonic orders ($\ell$) of 2-4,
while the Si-shell shows broad spectra on smaller scales of $\ell\approx30-40$.
Both convective regions show an increase in power at $\ell=5$ near collapse.
We show that the 1D \texttt{MESA} model agrees with the convective
velocity profile and speeds of the Si-shell when compared to our highest resolution 3D model. 
However, in the O-shell region, we find that \texttt{MESA} predicts speeds approximately \emph{four} 
times slower than all of our 3D models suggest. All eight of the multi-dimensional stellar models 
considered in this work are publicly available.
\end{abstract} 

\keywords{convection --- hydrodynamics --- stars: evolution --- stars: interiors --- supernovae: general}

\section{Introduction} 
\label{sec:intro}

Stars with an initial zero age main-sequence (ZAMS) mass of greater than approximately 
8-10 \msun may end their lives via core-collapse supernova (CCSN) explosions 
\citep{janka_2012_aa,farmer_2015_aa,woosley_2015_aa,woosley_2002_aa}. Core-collapse supernova explosions  
facilitate the evolution of chemical elements throughout galaxies 
\citep{timmes_1995_aa,pignatari_2016_aa,cote_2017_aa}, 
produce stellar mass compact object systems 
\citep{ozel_2012_aa,sukhbold_2016_aa, couch_2019_aa}, 
and provide critical feedback to galaxy and star formation
\citep{hopkins_2011_aa,botticella_2012_aa,su_2018_aa}.
Hydrodynamic simulations of CCSNe have helped inform 
our understanding of all of these aspects, beyond that which can be inferred directly from current observations.

\added{CCSN simulations now include 3D hydrodynamics as well as more accurate treatments of key physical aspects of the problem.}
Many advances have been made to produce such simulations, 
such as the inclusion of \added{two-moment} neutrino transport schemes 
\citep[e.g.,][]{hanke:2013, lentz:2015, oconnor_2018_ab, glas:2019, vartanyan:2019},
a general relativistic treatment for gravity as opposed to a Newtonian approach \citep{roberts_2016_aa, muller:2017},
and spatial resolutions that allow us to accurately capture the Reynolds stress \citep{radice:2016, nagakura_2019_aa}, 
a key component in the dynamics of the shock. Despite these advances, the vast majority 
of these simulations rely on one dimensional (1D) initial conditions for the progenitor star. These progenitors 
are typically produced using stellar evolution codes where convection is treated using mixing length theory 
(MLT) \citep{bohm_1958_aa,cox_1968_aa}. MLT has been shown to accurately represent convection in 
1D models when calibrated to radiation hydrodynamic simulations of surface convection in 
the Sun \citep{trampedach_2014_aa}. Despite the utility of MLT in 1D stellar models, multidimensional 
effects in the late stages of nuclear burning in the life of a massive star can lead to initial conditions 
that differ significantly from what 1D stellar evolution models suggest 
\citep{arnett_2009_aa,arnett_2011_ab,viallet_2013_aa}.

\citet{couch_2013_aa} investigated the impact of asphericity in the progenitor star on the explosion of a 
15 \msun stellar model that has been investigated in detail \citep{woosley_2007_aa}. Motivated by 
results of multidimensional shell burning in massive stars \citep{meakin_2007_aa,arnett_2011_ab}, they implemented 
velocity perturbations within the Si-shell to assess the impact on the explosion dynamics. They found that
the models with the velocity perturbations either exploded successfully or evolved closer to explosion where 
the models without the perturbations failed to successfully revive the stalled shock. The non-radial velocity 
perturbations resulted in stronger convection and turbulence in the gain layer.
These motions play a significant role in contributing to the turbulent pressure and dissipation which can  
supplement the thermal pressure behind the shock and, thus, enable explosions at lower effective neutrino 
heating rates \citep{couch_2015_ab,mabanta_2018_aa}. \added{\citet{couch_2013_aa} compared models 
with and without perturbations and with either fiducial or slightly enhanced neutrino heating. Their 3D model 
\emph{without} initial perturbations but slightly enhanced heating (2\% larger than fiducial)
followed a similar trajectory as the perturbed model (peak perturbation Mach number of $\mathcal{M}_{\rm{pert}}=0.2$) 
with no enhanced heating. However, neither of these models were able to revive the 
stalled shock and both resulted in a failed explosion.} These results suggested 
that the multidimensional structure of the progenitor star can provide a favorable impact 
on the likelihood for explosion \added{by increasing the total stress, both thermal and turbulent, behind the shock. Without perturbations, the progenitor star used by \citet{couch_2013_aa} required more neutrino heating (5\% larger than fiducial) to achieve shock revival \citep{couch_2014_aa}.}

One of the first efforts to produce multidimensional CCSN progenitors began with 
the seminal work of \citet{arnett_1994_aa}. They performed hydrodynamic O-shell burning 
simulations in a two-dimensional wedge using an approximate 12 species network. In this work,
they found maximum flow speeds that approached $\approx$ 200 km s$^{-1}$ which induced density 
perturbations and also observed mixing beyond stable boundaries. Much later, \citet{meakin_2007_aa}
presented the first results of O-shell burning in 3D. They evolved a 3D wedge encompassing the 
O-shell burning region using the \texttt{PROMPI} code for a total of about eight turnover timescales.
In comparing the 3D model to a similar 2D model they found flow speeds in the 2D model were significantly
larger and also found that the interaction between the convectively stable layers with large convective plums 
can facilitate the generation of waves. The combined efforts of these previous works all suggested the 
need for further investigation into the role of the late time properties of CCSN progenitors near collapse. 

Building upon previous work, \citet{couch_2015_aa} (hereafter C15) presented the first three-dimensional (3D) 
simulation of iron core collapse in a 15\msun star. They evolved the model in 3D assuming octant symmetry and an
approximate 21 isotope network for a total of $\approx$160 s up to the point of gravitational instability and iron core collapse. Their simulation captured $\sim$8 convective turnovers in the Si-burning shell region
with speeds in the Si-shell region on the order of several hundred km s$^{-1}$ and significant non-radial kinetic energy. 
They then followed the 3D progenitor model through core collapse and bounce to explosion using similar 
methods as in \citet{couch_2014_aa}, i.e. parameterized deleptonization, multispecies neutrino leakage scheme, 
and Newtonian gravity. When comparing the explosion of the 3D progenitor model to the angle-average of 
the same model, \added{they find that the turbulent kinetic energy spectrum ahead of the shock front (in the accretion 
flow) was more than an order-of-magnitude larger for the 3D case during accretion of the Si-shell, around post-bounce times 
of $t_{\rm{pb}}=125\pm25$ ms. The more turbulent accretion flow led to enhanced total turbulent 
kinetic energy in the gain region by up to almost a factor of two for the 3D initial conditions than the 
angle-averaged model, with most of the turbulent kinetic energy residing at scales of $\ell \approx6-10$, where $\ell$ is the principle spherical harmonic order. These differences resulted in an overall more
rapidly expanding shock radius and a diagnostic explosion energy approximately a factor of two larger than the 
1D initial model. However,} the model presented in that work suffered from approximations 
that may have affected the results. The main issues were the use of octant symmetry and the modification of the
electron capture rates used in the simulation. The first of these approximations can lead to a suppression of 
perturbations of very large scales while the second can lead to larger convective speeds within the Si-shell 
region due to the rapid artificial contraction fo the iron core. 

\citet{muller_2016_aa} aimed to address these issues by conducting a full 4$\pi$ 3D simulation of
O-shell burning in a 18\msun progenitor star. Using the \texttt{Prometheus} hydrodynamics code they 
evolved the model in 3D for $\sim$294 s up to the point of iron core collapse. They alleviate the use 
of enhanced electron capture rates by imposing an inner boundary condition that follows the radial trajectory 
of the outer edge of the Si-shell according to the 1D initial model generated by the \texttt{Kepler} stellar evolution code. In their
simulation, they capture approximately 9 turnover timescales in the O-shell finding Mach numbers that reach 
values of $\sim$0.1 as well a large scale $l=2$ mode that emerges near the point of collapse. Their results 
build on those of C15 with further evidence suggesting the need for full 4$\pi$ simulations. 
Recently, \citep{yadav_2019_aa} presented a 4$\pi$ 3D simulation of O-shell burning where a violent 
merging of the O/Si interface and Ne layer merged prior to gravitational collapse. This simulation 
of an 18.88 \msun progenitor for 7 minutes captured the mixing that occurred after the merging and 
was found to lead to Mach numbers of $\sim$0.13 near collapse. All of these efforts suggest that 3D progenitor structure increases the likelihood for explosion of massive stars by the delayed neutrino heating mechanism.

In this paper, we present 1D, 2D, and 3D hydrodynamical simulations of Si- and O- shell burning of a 
15 \msun progenitor star for the final $\sim$424 seconds of its life up to the point of gravitational instability and
iron core collapse. \added{Using these models, we: 
1) provide a detailed description of the convective regions in the Si- and O- burning shells,
2) estimate key stellar evolution parameters that may impact the explosion properties of CCSNe and compare them to 
their 1D counterparts, and 
3) study how the properties of these models may depend on resolution,  dimensionality/symmetry, and initial perturbations.}
\added{In order to self-consistently simulate secular core contraction and ultimate collapse, we include the iron core} and both 
Si- and O-shell regions in our simulation. To address the dependence of resolution, \added{dimensionality, and 
symmetry} on our results we consider 2D and 3D models at varying finest resolution and cylindrical versus
octant symmetry, respectively. We improve on the work of C15 by alleviating the use of accelerated electron capture rates, 
evolving two 3D simulations \added{that cover the full solid angle (4$\pi$ steradian)} rather than octant symmetry, 
and significantly increase the timescale of the simulation. \added{Lastly, to assess the impact of initial conditions, 
our two 4$\pi$ 3D models differ only in the initialization of the velocity field.} Using these models, we characterize
 the qualitative properties of the flow 
amongst different parameter choices while also comparing to the predicted properties of the 
1D input model for our simulations. We present a detailed analysis of the convective properties of the 4$\pi$ 3D models and discuss the implications for the 3D state of CCSN progenitors at collapse. This paper is organized as follows. In Section~\ref{sec:methods} we discuss our computational methods and input physics, in Section~\ref{sec:results} we 
present the results of 
our 2- and 3D hydrodynamical \FLASH simulations, Section~\ref{sec:discussion} summarizes our results 
and compares them to previous efforts.

\section{Methods and Computational Setup} 
\label{sec:methods}
Our methods follow closely those used in C15 in that we evolve a 1D 
spherically symmetric stellar evolution model using \MESA and, at a point near iron core collapse,
map the model into the \FLASH simulation framework and continue the evolution in multi-D to 
collapse. In the following subsections, we will describe these steps in detail. 

\subsection{1D \MESA Stellar Evolution Model}

We evolve a 15 \msun solar metallicity stellar model using the 
open-source stellar evolution toolkit, Modules for Experiments in Stellar Astrophysics (\texttt{MESA})
\citep{paxton_2011_aa,paxton_2013_aa,paxton_2015_aa,paxton_2018_aa,paxton_2019_aa}.
The model is evolved from the pre-main sequence to a time approximately 424 seconds before 
iron core collapse, defined by \MESA as the time when any location of the iron core reaches
an infall velocity of greater than 1000 km s$^{-1}$. We use temporal and spatial parameters similar to those 
used in \citet{farmer_2016_aa} and \citet{fields_2018_aa}. These parameters result in timesteps on the 
order of $\bar{\delta t}\approx$ 41 kyr during the main sequence, $\bar{\delta t}\approx$ 24 yr during 
core carbon burning, and  $\bar{\delta t}\approx$ 19 sec when the model is stopped. At the point when 
the model is stopped, the model has 3611 cells with an enforced maximum $\bar{\delta m}\approx$ 
0.01 \Msun. The \MESA model uses an $\alpha$-chain network that follows 21 isotopes from 
$^{1}$H to $^{56}$Cr. This network is chosen for its computational efficiency and to match the network 
currently implemented in \FLASH and used in \citep{couch_2015_aa}. This approximate network aims 
to capture important aspects of pre-supernova evolution of massive star models which include reactions between
heavy ions and iron-group photodisintegration and similar approaches have been used in many previous studies 
\citep{heger_2000_aa,heger_2010_aa}. We include mass loss using the `\texttt{Dutch}` wind scheme 
with an efficiency value of 0.8. Mixing processes due to convective overshoot, thermohaline, and semi-convection 
are considered with values from \citet{fields_2018_aa}. We do \added{not} include rotation and magnetic fields in this model. 

\begin{figure}[!htb]
         \centering  
        \begin{subfigure}{
                \includegraphics[width=0.47\textwidth]{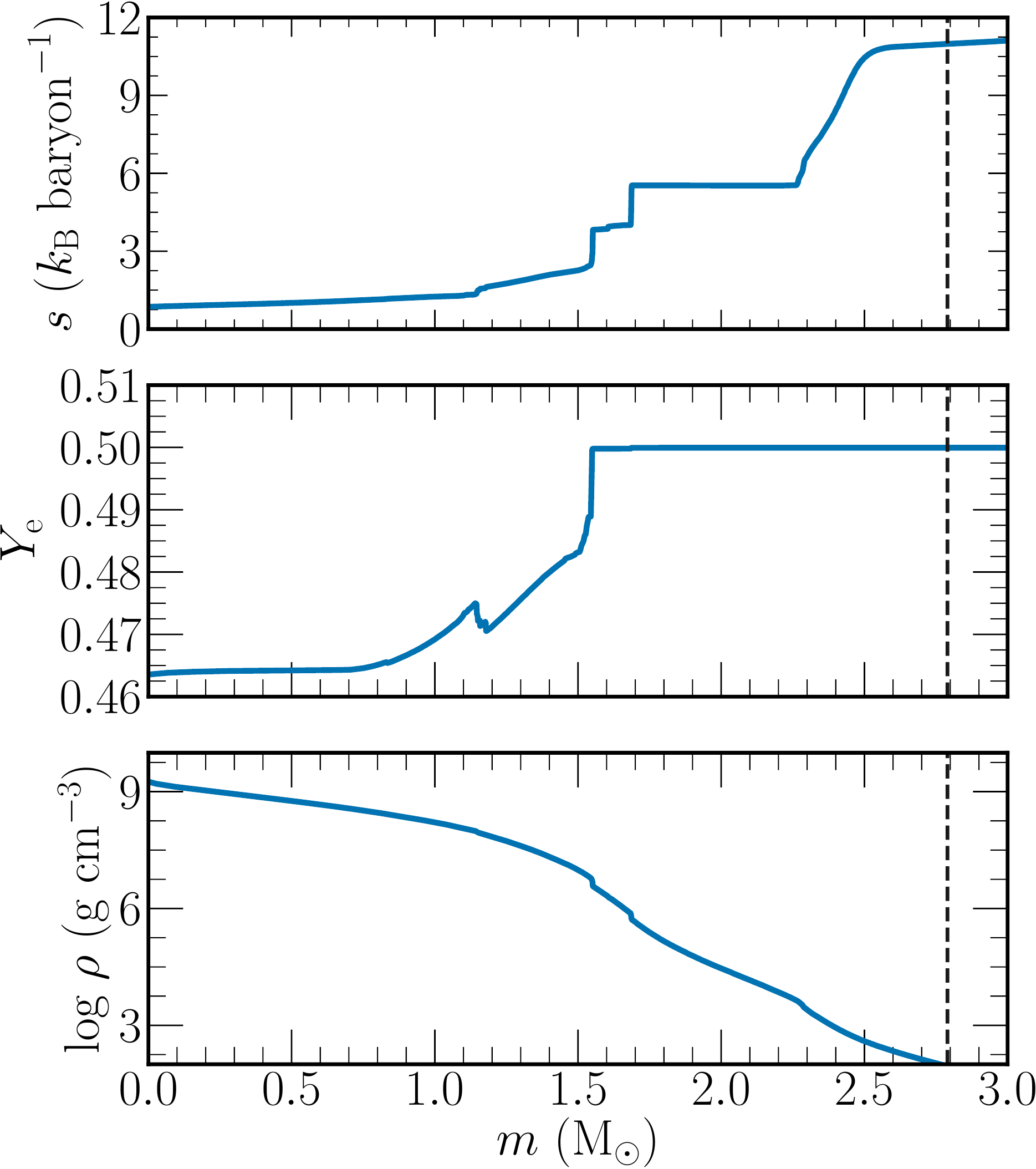}}
        \end{subfigure}
        \caption{
Specific entropy, electron fraction, and mass density profiles as
function of stellar mass for the 1D \MESA model at the time of 
mapping into \FLASH. The dashed black vertical line 
denotes the edge of the domain used in the \FLASH simulations.
        }\label{fig:1d_mesa_profiles}
\end{figure}

In Figure~\ref{fig:1d_mesa_profiles} we show the specific entropy, electron fraction, and mass density 
profiles  as a function of mass coordinate from the \MESA model at the point which it is 
mapped into \FLASH. The dashed black line denotes the edge of 
the domain considered in the \FLASH simulations. 
Overall, the model used in this work has a similar structure to the progenitor used in C15 except for the case
of the lower central electron fraction in the core. This difference is partially due to the different network used in 
C15, a basic 8 isotope network that was automatically extended during evolution compared to our static 21
isotope network used in the \MESA model for this work. The input \MESA model in C15 also had a slightly
smaller initial iron core mass, $1.3$ \msun, than the model considered here.

\begin{figure}[!htb]
         \centering  
        \begin{subfigure}{
                \includegraphics[width=0.47\textwidth]{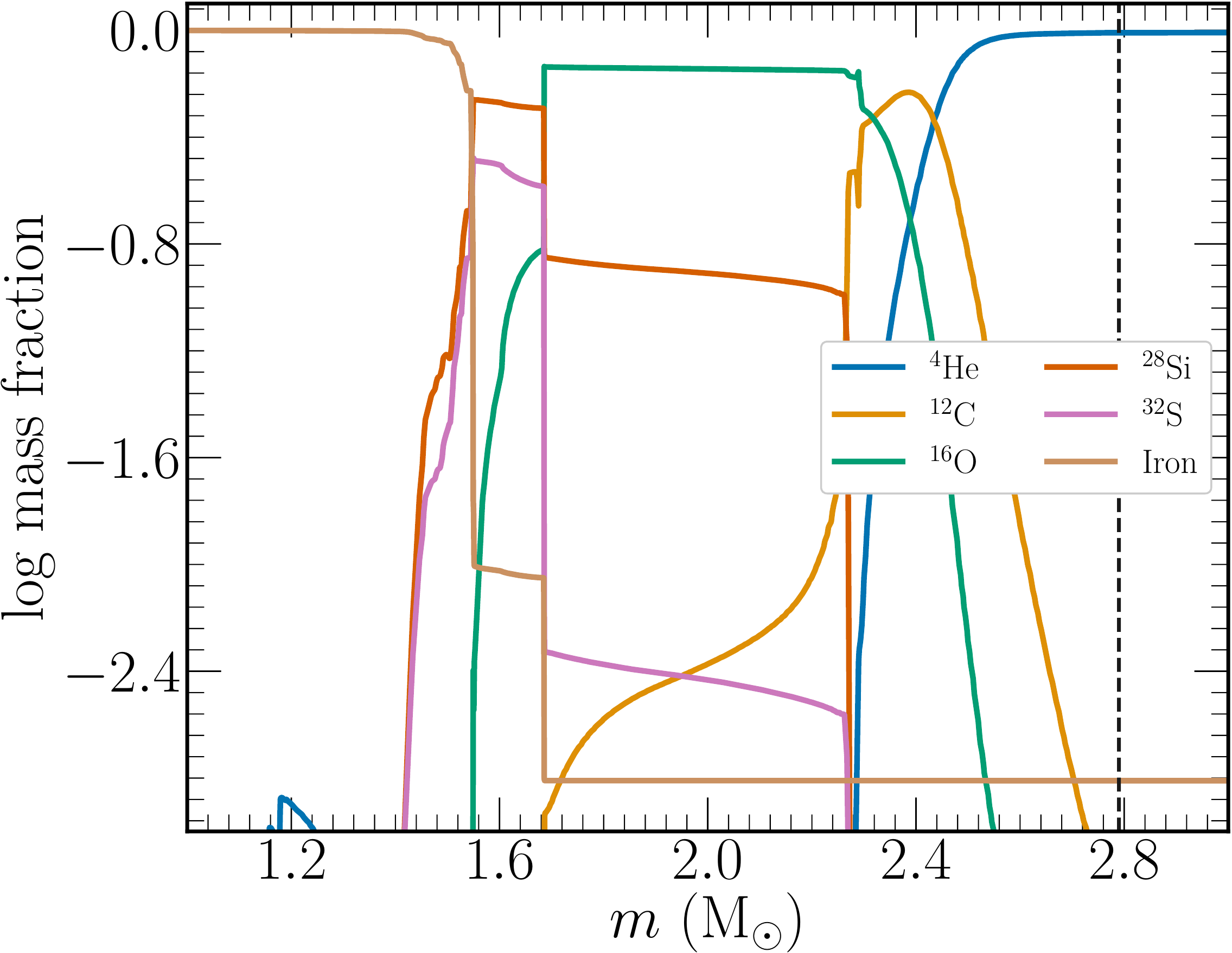}}
        \end{subfigure}
        \caption{
Mass fractions of the most abundant isotopes as a 
function of stellar mass for the 1D \MESA model at time of mapping into \FLASH. The label `Iron' denotes the 
sum of mass fractions of $^{52,54,56}$Fe isotopes. The dashed black vertical line denotes the edge of the 
domain in \FLASH.
        }\label{fig:1d_composition}
\end{figure}

\begin{figure*}[!htb]
         \centering  
        \begin{subfigure}{
                \includegraphics[width=0.482\textwidth]{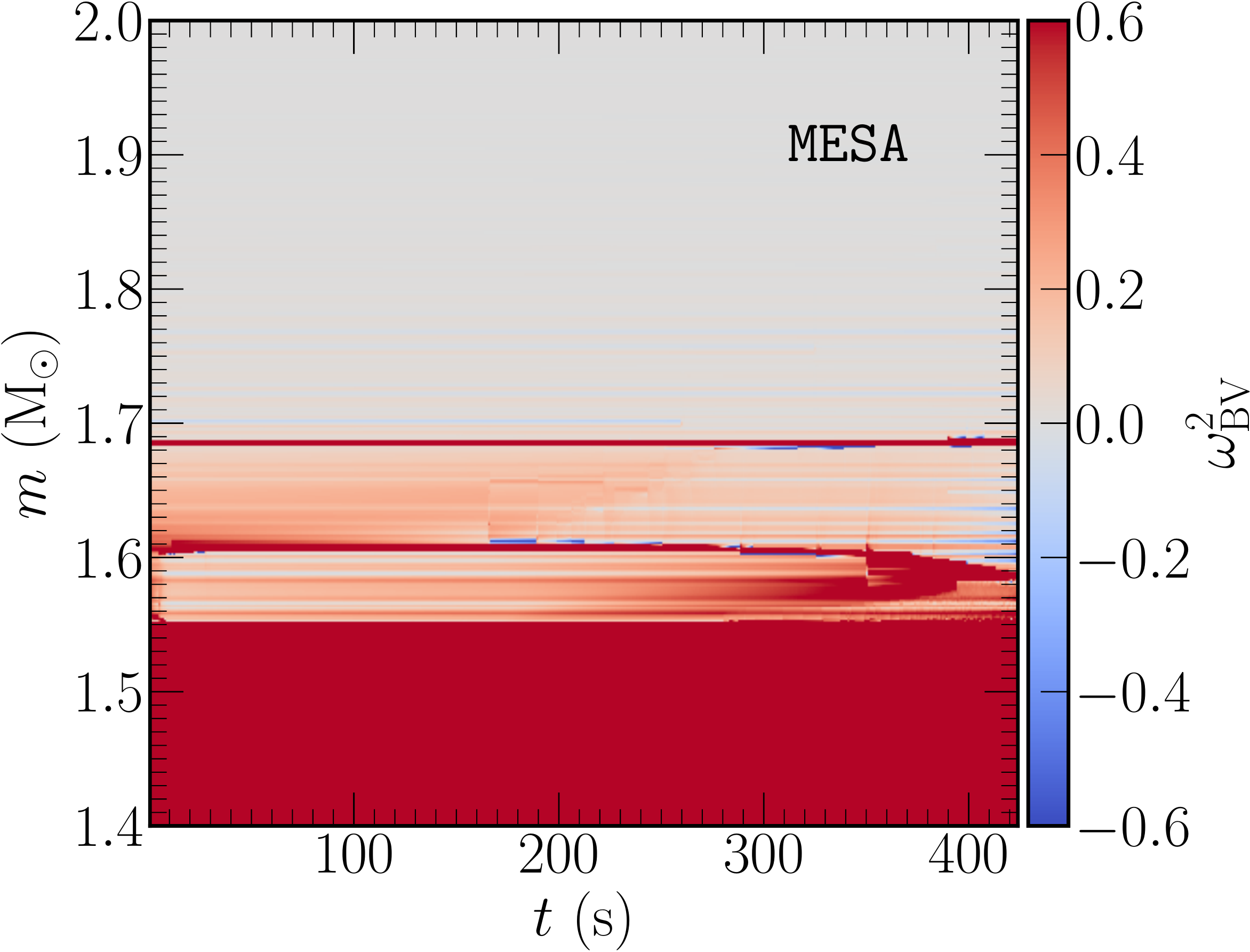}}
        \end{subfigure}
        \begin{subfigure}{
                \includegraphics[width=0.47\textwidth]{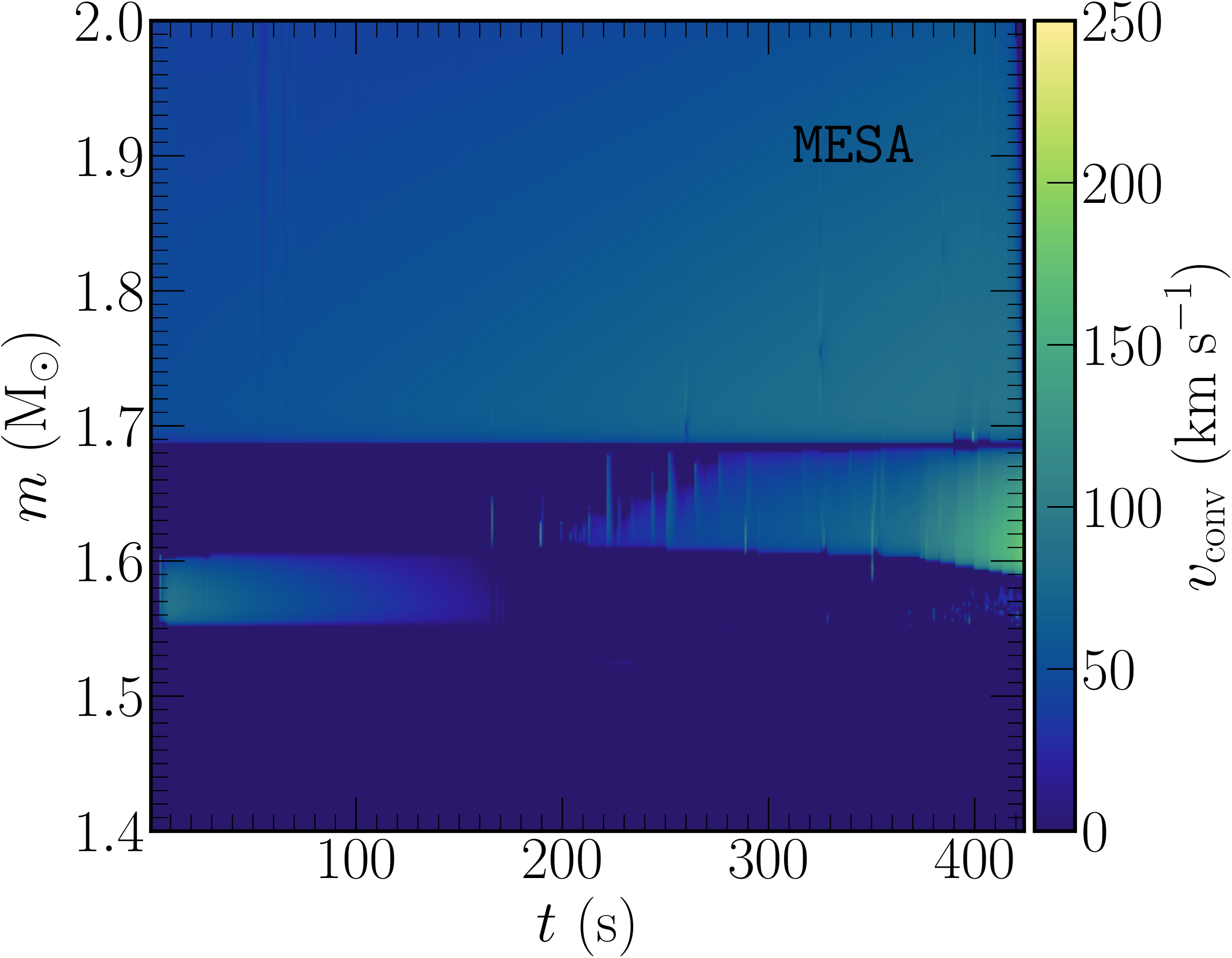}}
        \end{subfigure}
        \caption{
Time evolution of the 1D profile data for the \MESA model. The 
(left) subplot shows the \BV\ frequency while the (right) shows the convective velocity speeds according to MLT. 
The red regions denote regions that are stable against convection while gray and blue regions 
show regions that are unstable to convection according to \texttt{MESA}.
        }\label{fig:1d_conv_time_evol}
\end{figure*}

Figure~\ref{fig:1d_composition} shows the mass fraction profiles for some of the most abundant isotopes for the input 
\MESA model. The label `Iron' denotes the sum of mass fractions of $^{52,54,56}$Fe isotopes. 
At the point of mapping the stellar model has an iron core mass of approximately 1.44 $M_{\odot}$. 
The Si-shell region is located at a specific mass coordinate of $m\approx 1.53-1.68$ \msun while
the O-shell region extends from the edge of the Si-shell out to mass coordinate of $m\approx 2.26$ \msun.
Figure~\ref{fig:1d_conv_time_evol} shows the time evolution of the \BV\ frequency (left) and 
convective velocity speeds \added{(right)} as a function of mass coordinate as predicted by \MESA for the 1D model evolved 
from the point at which it is mapped into \texttt{FLASH} until core collapse. The 1D model predicts convective 
speeds in the O-shell regions with a peak of approximately 100 km s$^{-1}$ up to the point of collapse. In the Si-shell 
region, only the inner most region is convectively active with speeds on the order of those in the O-shell.
At a time of $t\approx200$ s, the innermost Si-shell burning convective region ceases and convective 
proceeds instead at a further mass coordinate of $m\approx 1.60-1.68$ \msun with speeds increasing to
values greater than in the O-shell near collapse at $\approx 160 $ km s$^{-1}$.

\subsection{2- and 3D \FLASH Stellar Evolution Model}
\label{sec:methods_mD}

\subsubsection{Overview}
We perform a total of \added{eight} multidimensional stellar evolution models at various resolutions and symmetries. 
All models are evolved using the 
\FLASH simulation framework \citep{fryxell_2000_aa,dubey_2009_aa}. Similar to \citet{couch_2015_aa}, we utilize the 
``Helmholtz'' EoS \citep{timmes_2000_aa} and the same 21 isotope network but with an improvement to the 
weak reaction rate used for electron capture onto $^{56}$Ni. The original network used tabulated rates from 
\citet{mazurek_1974_aa} while the updated rates were adopted from \citet{langanke_2000_aa}. 
The new rates are enhanced by close to a factor of 5-10 alleviating any need to artificially enhance 
the total electron capture rates in the models presented in this work and are also in agreement with the
table values used in \MESA.

\subsubsection{Hydrodynamics, Gravity, and Domain}
The equations of compressible hydrodynamics are solved using \texttt{FLASH}'s directionally unsplit piecewise 
parabolic method (PPM) \citep{lee_2008_aa} and HLLC Riemann solvers \citep{toro_1999_aa} with a 
Courant factor of 0.8. Self-gravity is solved 
assuming a spherically symmetric (monopole) gravitational potential approximation \citep{couch_2013_ab}. 
Our computational domain extends to 10$^{10}$ cm from the origin in each dimension for both the 2D and 3D models. 
Four 2D models are evolved with varying levels of finest grid spacing resolution of 8,16, 24, and 32 km.
The 2D models use cylindrical geometry with symmetry about the azimuthal direction. 
\added{Four} 3D models are evolved: two assuming octant symmetry with two different values of 
finest grid resolution, 16 and 32 km, and two full $4\pi$ 3D models at 32 km finest grid resolution, \added{one 
with an initialized velocity field and one without}. All 3D 
models use Cartesian coordinates. The models are labeled according to their dimensionality and finest grid 
spacing, and in the case of 3D according to the use of octant symmetry or not. For example, the 3D 
32 km octant model is labelled \texttt{3DOct32km} for ease of model identification throughout the 
remainder of this paper. \added{The initially perturbed $4\pi$ 3D model is labeled $\texttt{3D32kmPert}$}. 

\added{The velocity 
field initialization for the perturbed model follows the methods used in \citet{muller_2015_ab} and extended to 3D in \citet{oconnor_2018_aa}. 
The method introduces solenoidal velocity perturbations to the $v_{r}$ and $v_{\theta}$ components using spherical harmonics and sinusoidal radial dependence. We use the convective velocity profile 
of the 1D \MESA model at the time of mapping to inform our choices of parameters to initialize the velocity field. 
We take the innermost convective Si-shell region to be from $r_{\rm{1,min}}\approx$ 2220 km to 
$r_{\rm{1,max}}\approx$ 2740 km, and choose 
spherical harmonic coefficients to be $\ell=5,m=1$ with a single radial sinusoid ($n=1$). The second convective Si-shell region, 
at $r_{\rm{2,min}}\approx$ 2740 km to $r_{\rm{2,min}}\approx$ 3760 km, uses the same
spherical harmonic and radial numbers as the first; $n=1,\ell=5,m=1$. 
Lastly, the O-shell region, taken to be $r_{\rm{3,min}}\approx$ 3770 km to $r_{\rm{3,max}}\approx$ 26,620 km, 
assumes an initially larger scale flow with quantum numbers of $n=1,\ell=5,m=3$. }

\added{
All velocity amplitude scaling factors ($C$) for the initial perturbations were chosen to match approximately 1$\%$ (in the Si-shell regions) and 5$\%$ (in the O-shell region) 
of the convective velocity speed as predicted by the \MESA model at time of mapping. The resulting perturbations 
led to an angle-averaged Mach number of $\mathcal{M}\approx 2\times10^{-3}$ and $\mathcal{M}\approx 1.3\times10^{-2}$ 
in the Si-shell and O-shell regions, respectively. For comparison, the scaling factors used in 
\citep{oconnor_2018_aa} to replicate pre-collapse perturbations were on the order of $\mathcal{O}\sim10^{3}$ 
larger than values used here. In terms of the total initial kinetic energy the perturbed model begins with a value of 
$E_{\rm{kin.}} \approx 5.8\times10^{45}$ erg, approximately 4\% of the peak total kinetic energy observed at later times.}

All models utilize adaptive 
mesh refinement (AMR) with up to eight levels of refinement. To give an example of the grid structure for 
our models we consider the refinement boundaries for the \texttt{2D8km} and \texttt{2D32km} models. In \added{both of these models,} the entire Si-shell region has a grid resolution of 32 km with effective angular resolution of 
$\approx$ 0.9$^{\circ}$ to 0.5$^{\circ}$ at the base (\added{$\approx$} 2000 km) and edge of the shell (\added{$\approx$} 
3500 km), respectively. 
\added{The resolution within the Si-shell region for these model corresponds to an average of $\approx$ 5\%  
of the local pressure scale height, $H_{\rm{p}}$.}
The grid resolution in the 2D models are representative of the respective 3D models as well. 
The two finer 
resolution levels of \added{the \texttt{2D8km}} model are situated within the iron core with the second highest level of refinement at the 
base of the Si-shell region and down to a radius of 1000 km. \added{In this region, the \texttt{2D8km} and \texttt{2D32km}
grid resolutions provide a value of $\approx$ 6\% and $\approx$ 12\% of $H_{\rm{p}}$, respectively}. The O-shell region in 
\added{these models} is at 64 km \added{resolution} out to a radius of $\approx$ 6000 km then decreases to 128 km from 
there to 10,000 km. 
Within these two regions the effective angular resolution ranges from 0.73$^{\circ}$ to 0.61$^{\circ}$ \added{and
corresponds to an average of $\approx$ 0.04$H_{\rm{p}}$ and $\approx$ 0.05$H_{\rm{p}}$, respectively}. In the 
\texttt{2D32km}, the finest resolution level goes out to a radius of 3500 km, the approximate edge of the Si-shell, 
giving the model similar resolution in this and the O-shell to that of the \texttt{2D8km} without the two finer resolution 
levels within the iron core. \added{The location of refinement levels used is determined as a function of radius and chosen 
based on logarithmic changes in specific density, pressure, and velocity. The refinement levels are static throughout the 
simulation and chosen based on the input model.}

For the 3D octant symmetry planes and the 2D axis plane 
we use reflecting boundary conditions while the outer boundaries utilize a boundary condition that applies power-law extrapolations
of the velocity and density fields to approximate the roughly hydrostatic outer envelope of the stellar interior. The 4$\pi$ 3D models 
use the same 
custom boundary condition at all domain edges. To help \added{reduce} 
any \added{artificial} transients that occur from mapping \added{from a Lagrangian to an Eulerian code with different
grid resolution,} we use the approach of \citet{zingale_2002_aa}. \added{This approach takes the initial 1D \MESA model 
and maps it to a uniform grid with resolution four times finer than that used in our \FLASH grid. The density in the 
remapped model is then slightly modified while the pressure is held fixed to force the model into hydrostatic equilibrium 
(HSE), such that it satisfies 
\begin{equation}
\nabla P = \rho g~,
\end{equation}
in the absence of initial velocities. 
The procedure is then closed by calling the equation of state (EOS) for the new profile.}

\begin{figure*}[!htb]
         \centering
        \begin{subfigure}{
               \includegraphics[width=0.47\textwidth]{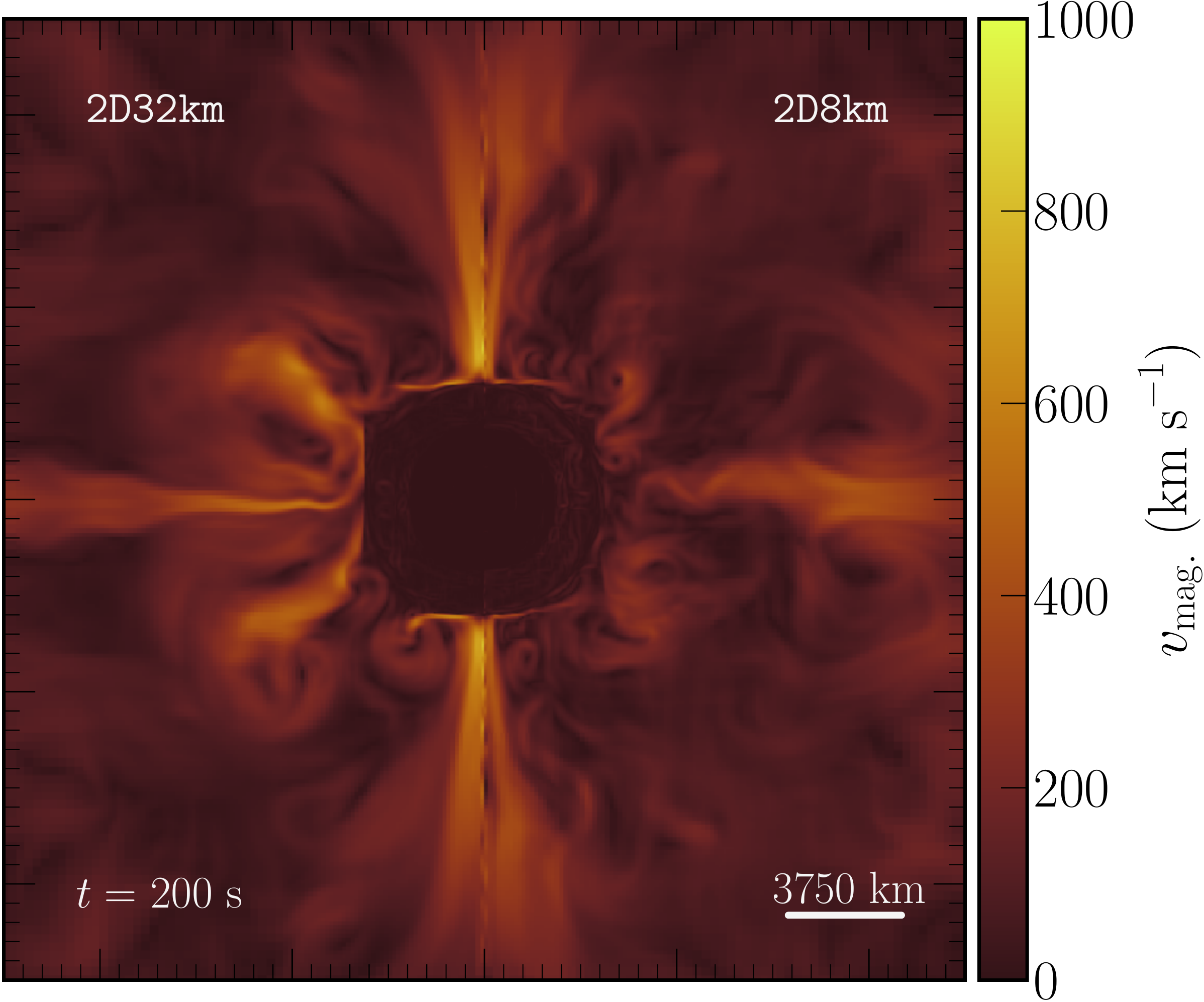}}
       \end{subfigure}
        \begin{subfigure}{
                \includegraphics[width=0.47\textwidth]{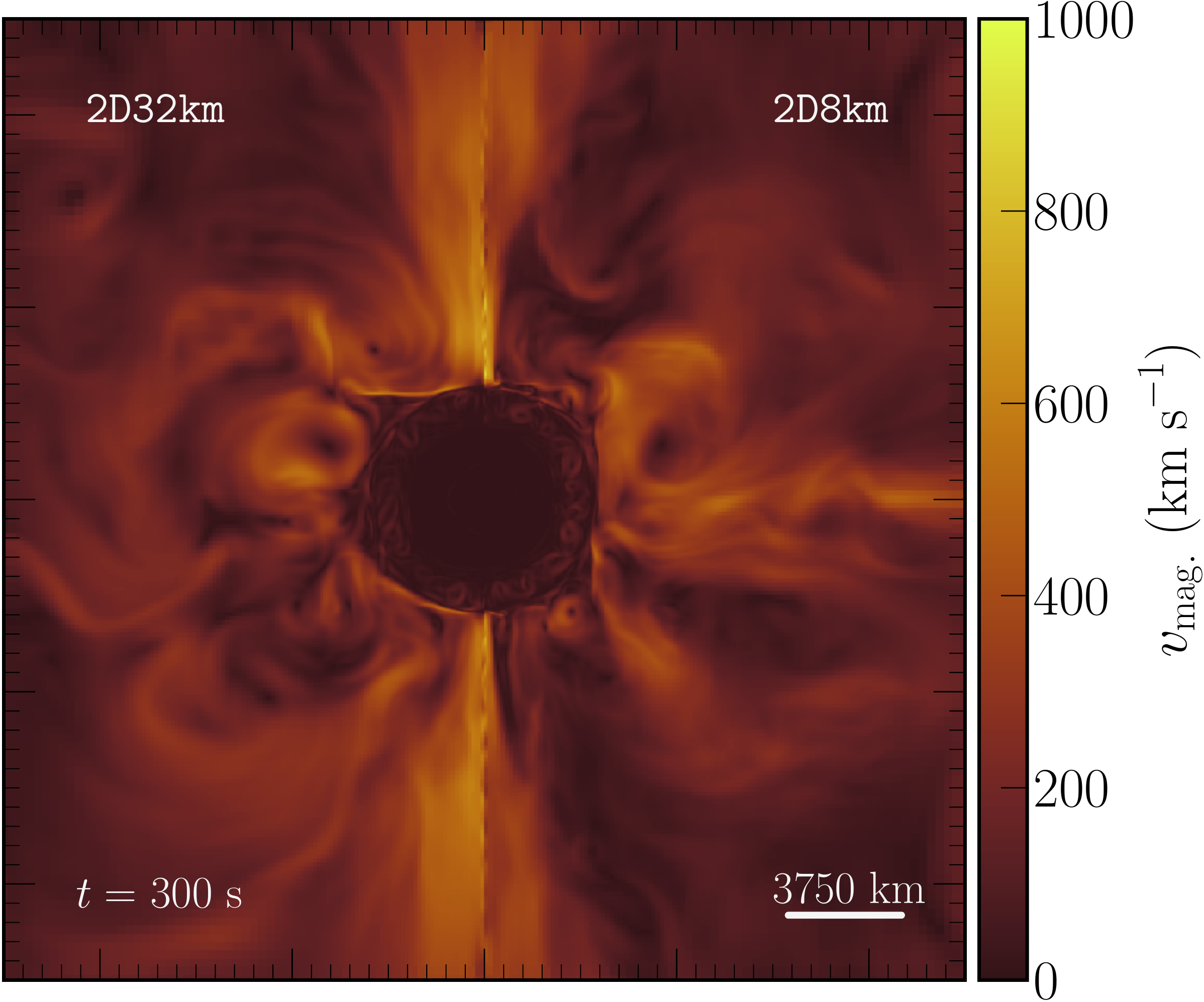}}
        \end{subfigure}          
        \begin{subfigure}{
                \includegraphics[width=0.47\textwidth]{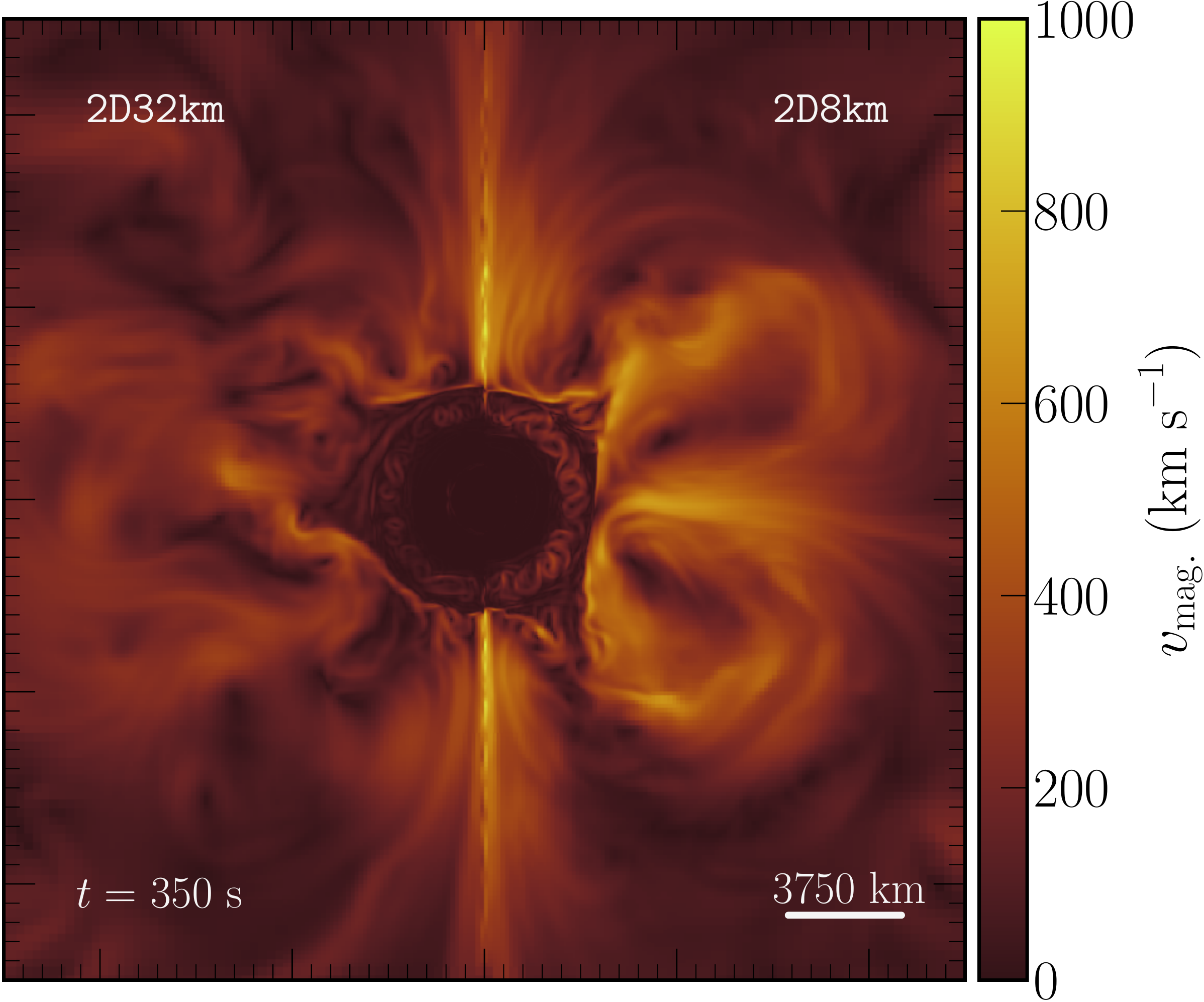}}
        \end{subfigure}
        \begin{subfigure}{
                \includegraphics[width=0.47\textwidth]{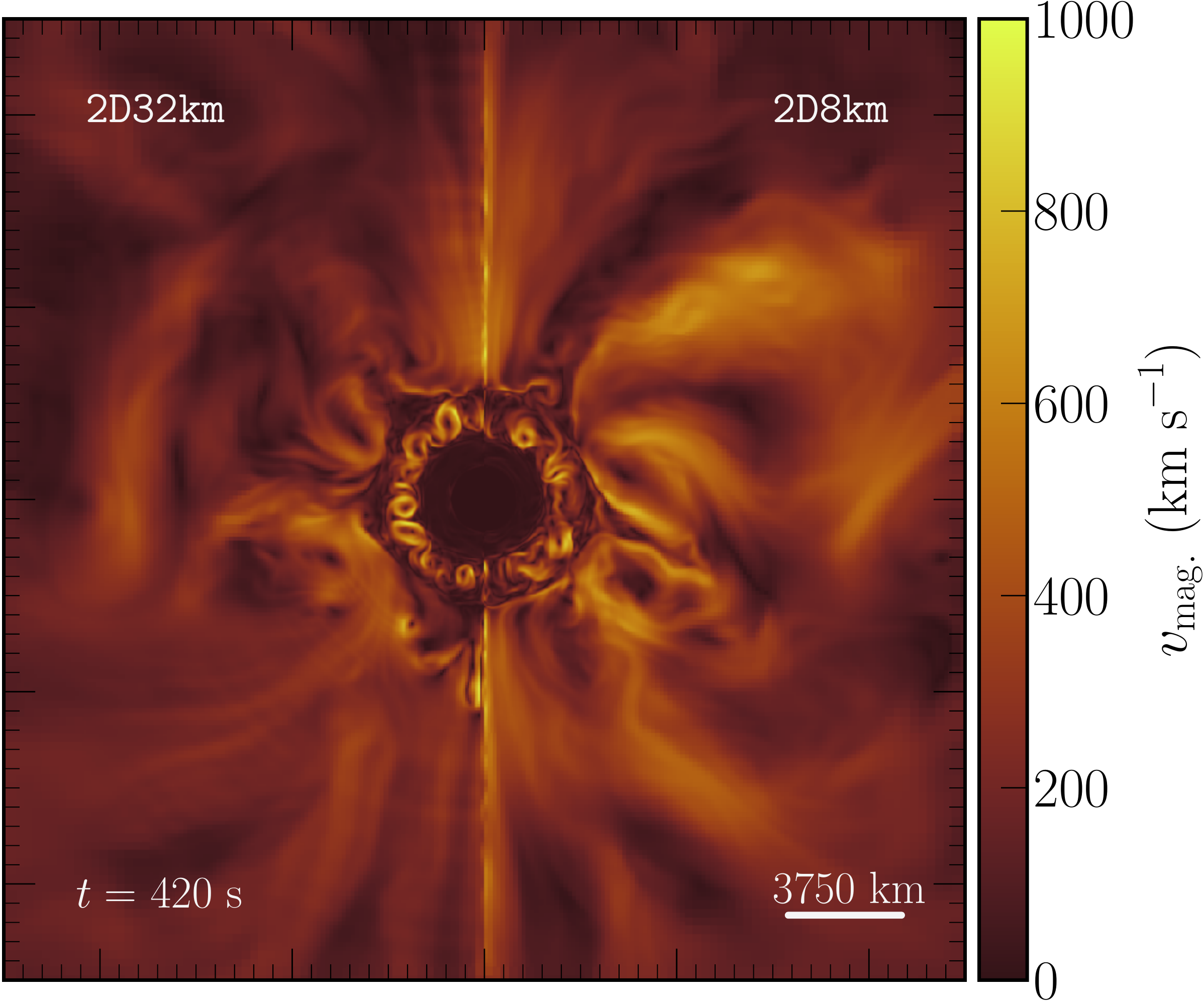}}
        \end{subfigure}
        \caption{
Slices of the the magnitude of the velocity field for the \texttt{2D32km} (left) and \texttt{2D8km} model (right) at four
different times. At $t= 200$ s, both models exhibit a ``square'' like imprint in the velocity field that is due to the stiff 
entropy barrier at the edge of the Si-shell region along with increased velocity speeds along the axis. These 
speeds are due to numerical artifacts common in 2D simulations assuming axisymmetry. The increased velocity along the axis causes outflows in the positive radial direction at the top and 
bottom of the Si-shell region. Beyond $t=200$ s, both models \added{become} characterized by large scale cyclones in the O-shell region with convective activity in the Si-shell having \added{three to five times} slower flow speeds.}
\label{fig:2d_si28_time_evol}
\end{figure*}

\subsubsection{Nuclear Burning}
During the growth of the iron core, thermal support is removed from the core due to photo-disintegration of iron nuclei
that cause the gas to cool. The subsequent contraction of the stellar core causes an increase in the rate of
electron captures onto protons therefore decreasing the pressure support contribution from electron 
degeneracy. This further reduction of pressure support in the core accelerates nuclear burning in the 
silicon burning shell leading to faster growth of the iron core. In the moments prior to iron core collapse
the core moves towards nuclear statistical equilibrium while the neutrino cooling and photo-disintegration rates
begin to dominate and lead to a negative specific nuclear energy generation rate, $\epsilon_{\rm{nuc}}$, in the 
inner core. Within this cooled material, the electron capture rates increase significantly and give rise to a positive 
specific $\epsilon_{\rm{nuc}}$ and cause the temperature to rise again. Within this region a numerical instability 
can occur for calculations which employ operator split burning and hydrodynamics \citep{timmes_2000_ab}. 
In order to avoid this instability from occurring in the models considered here, we place a limiter on the 
maximum timestep such that any change in the internal energy across 
all zones in the domain is limited to a maximum of a 1$\%$ per timestep. 

The large and opposite specific nuclear energy generation rates within the core can also lead to 
significant difficulties in solving the nuclear reaction network at a given timestep and lead to 
significant load imbalance of the computational workload per MPI rank. Zones within the 
core can take several hundred iterations to obtain a solution while outside of the iron core, 
a solution is found in a fraction of the time. In order to help circumvent these issues, we employ 
a moving ``inner boundary condition'' within the iron core, well below the region of interest for this paper. 
For all simulations considered here, the profiles of $\rho,Y_{\rm{e}}$, and $P$ in the inner 1000 km of 
the models are evolved according to the profile from the \MESA model at the corresponding time. 
A 2D table is constructed from the \MESA profile data from the point of mapping to \FLASH ($\approx 424$ s)
until iron core collapse. Four point Lagrange linear interpolation is then performed in time and 
radius to ensure accurate values are 
mapped for the \FLASH models, which take on the order of 100 timesteps for every \MESA timestep. 
This mapping effectively provides a time-dependent inner boundary condition that ensures the model
follows the central evolution of the \MESA model while still allowing us to capture the pertinent multi-D hydrodynamic 
behavior with \FLASH. 

For all models, we simulate $\approx$ 424 s of evolution prior to collapse capturing Si- and O-shell burning 
up to gravitational instability and iron core collapse. The full $4\pi$ 3D models had an approximate total of 
46M zones, took $\approx$ 0.6 M core hours on the \texttt{laconia} compute cluster at Michigan State University. 
All the multidimensional CCSN progenitor models considered in this work are available publicly at\dataset[10.5281/zenodo.3976246]{https://doi.org/10.5281/zenodo.3976246}.

\section{Multi-Dimensional Evolution to Iron Core-Collapse} 
\label{sec:results}

\subsection{Results from 2D Simulations}
\label{sec:2d}
We evolve a total of four 2D simulations at 8,16, 24, and 32 km finest grid spacing resolution. 
In the following subsection we consider the global properties of all of the 2D models, 
compare the lowest and highest resolution model, and
 consider in detail the convective properties of the highest resolution 2D model.
 
  \begin{figure}[!htb]
\centering{\includegraphics[width=1.0\columnwidth]{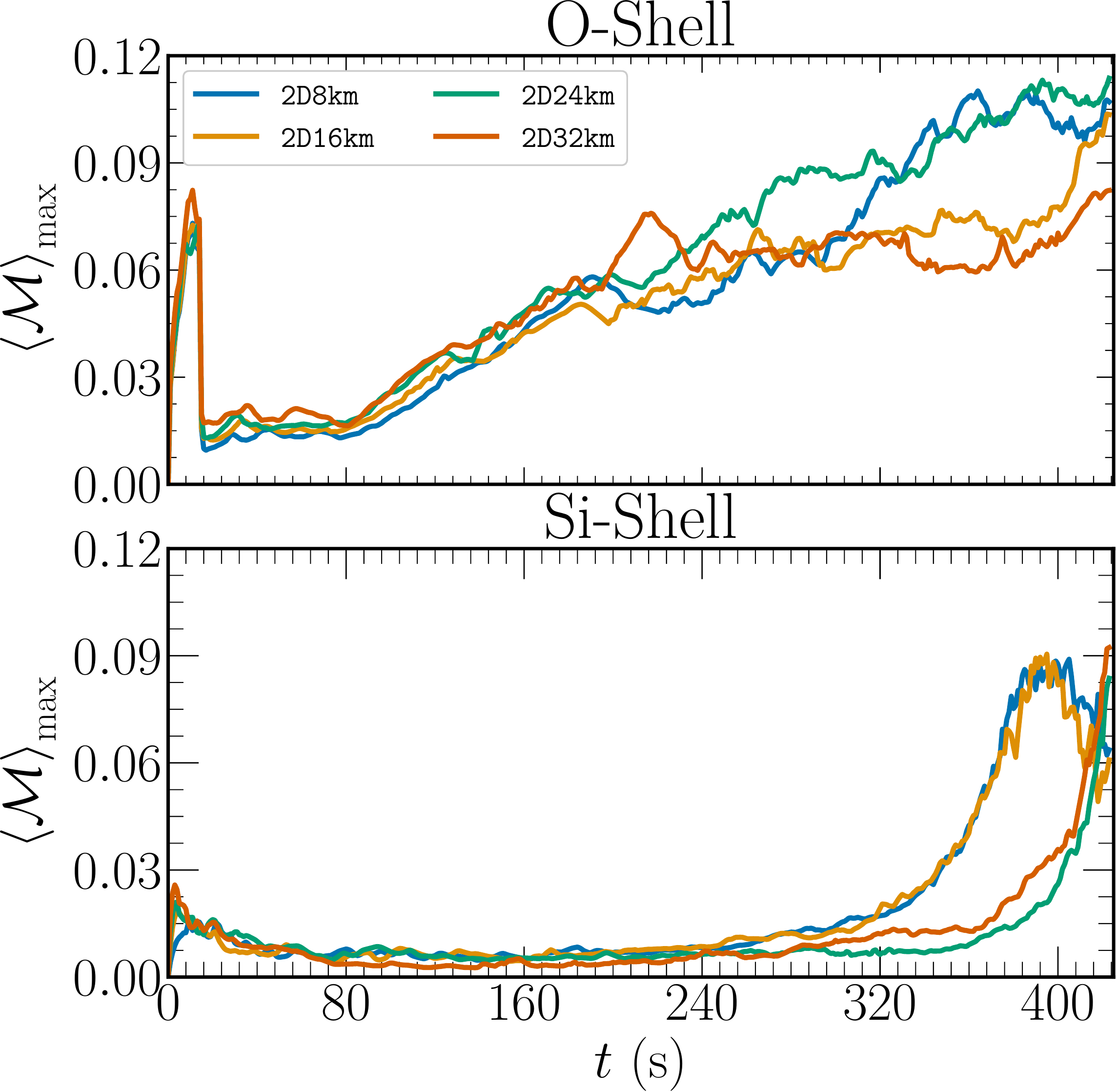}}
\caption{
Time evolution of the maximum angle-averaged Mach number within the Si- and O-shell for all 2D models. 
}\label{fig:2d_time_evol_max_mach_number_shells}
\end{figure}
 
The structure of the flow at large scale within the shell burning regions can have a significant impact on the 
CCSN explosion mechanism. Perturbations within these region can be amplified 
during collapse of the iron core and aid in the development of turbulence during explosion \citep{lai:2000, couch_2013_aa,couch_2015_ab}.
In Figure~\ref{fig:2d_si28_time_evol} we show slices of the magnitude of the velocity field for the
 \texttt{2D32km} (left \added{subplots}) and \texttt{2D8km} models (right \added{subplots}) at four different times. 
 At early times, we 
see convection developing in a similar matter for both the 8 km and 32 km models. Both models 
exhibit a ``square'' like imprint in the velocity field that is due to the stiff entropy barrier at the edge of the
Si-shell region interacting with the Cartesian-like 2D axisymmetric grid. 
\added{We also observe high-velocity flows restricted to the vicinity of the axis of symmetry, an artifact commonly observed in 2D axisymmetric simulations.}
\added{Around} $t=200$ s both models \added{become} characterized by large scale cyclones in the O-shell region \added{with convective speeds of $\approx100-120$ km s$^{-1}$, three to five times larger than in the Si-shell region.} 
At late times, beyond
$t=350$ s, the 8 km model appears to reach flow speeds in the Si-shell that are on the order of those observed 
in the O-shell. In the O-shell, the scale of the convection increases as cyclones on the order of 
$\approx$ 4000 km dominate the flow with the scale of convection within the Si-shell region being 
restricted by the width of the shell. Seconds prior to collapse, the 32 km model reaches Si-shell 
convective speeds that agree with the 8 km model.

\begin{figure}[!htb]
\centering{\includegraphics[width=1.0\columnwidth]{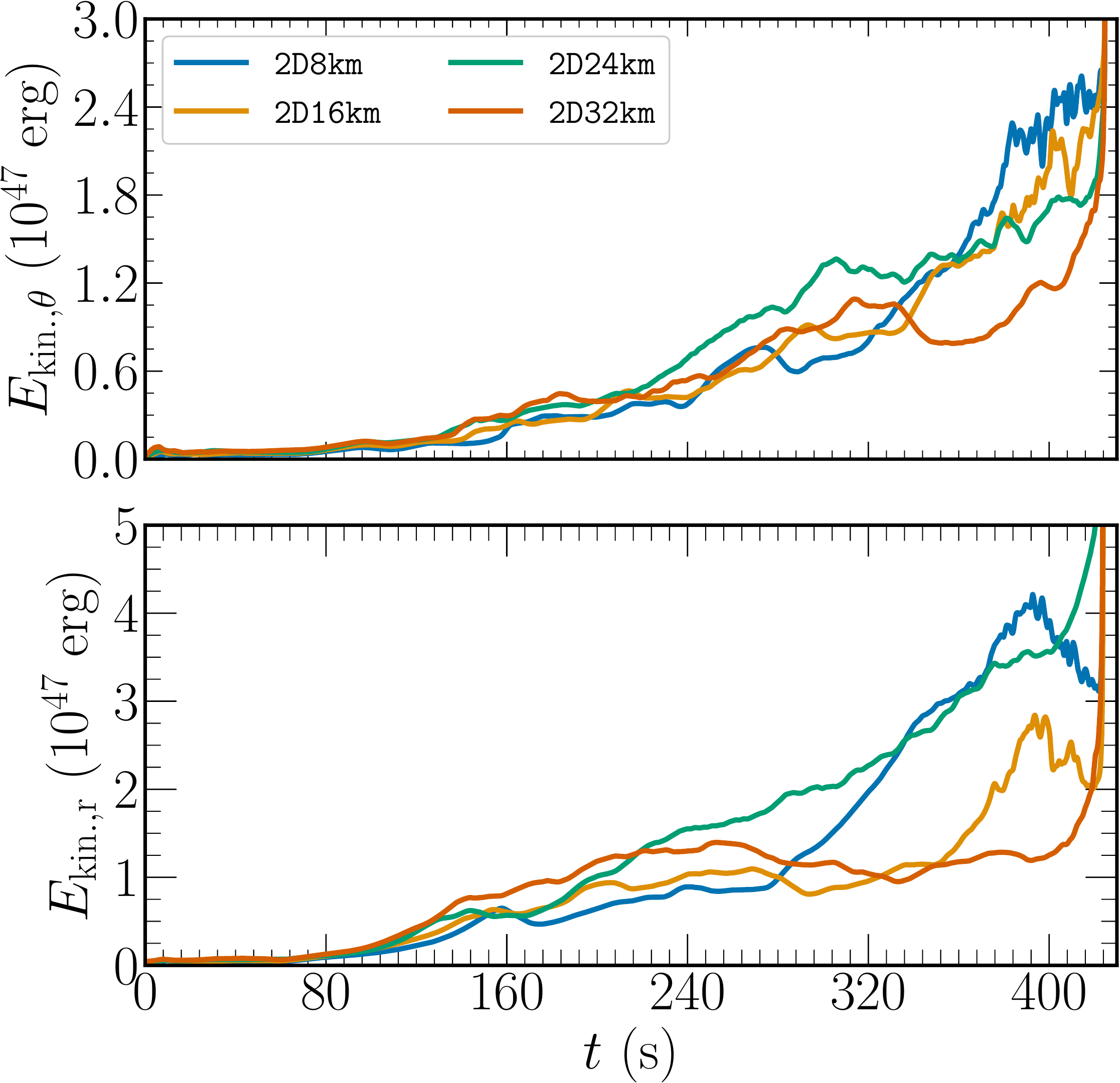}}
\caption{
Time evolution of the radial and non-radial kinetic energy for the four 2D models.
}\label{fig:2d_ekin_total_time_evol}
\end{figure}

To begin our assessment of the convective properties of the models, we compute 
the angle-averaged maximum Mach number, 
defined as,
\begin{equation}
\left < \mathcal{M} \right > = \left < \frac{\left | \bf{v} \right |}{c_{\rm{s}}} \right >~,
\end{equation}
where $\left | \textbf{v} \right |$ is the local magnitude of the velocity field and $c_{\rm{s}}$ the local sound speed, and the averaging is performed over solid angle. We 
also compute the Mach number within the Si-shell and within the O-shell to characterize their behavior independently. 
The shell region for silicon-28 is defined as the region where $X(^{28}\rm{Si}) > 0.2$ and $X(^{16}\rm{O}) < 0.2$ 
and for oxygen-16 $X(^{16}\rm{O}) > 0.2$ and $X(^{28}\rm{Si}) < 0.2$.

\begin{figure*}[!htb]
         \centering  
        \begin{subfigure}{
                \includegraphics[width=0.482\textwidth]{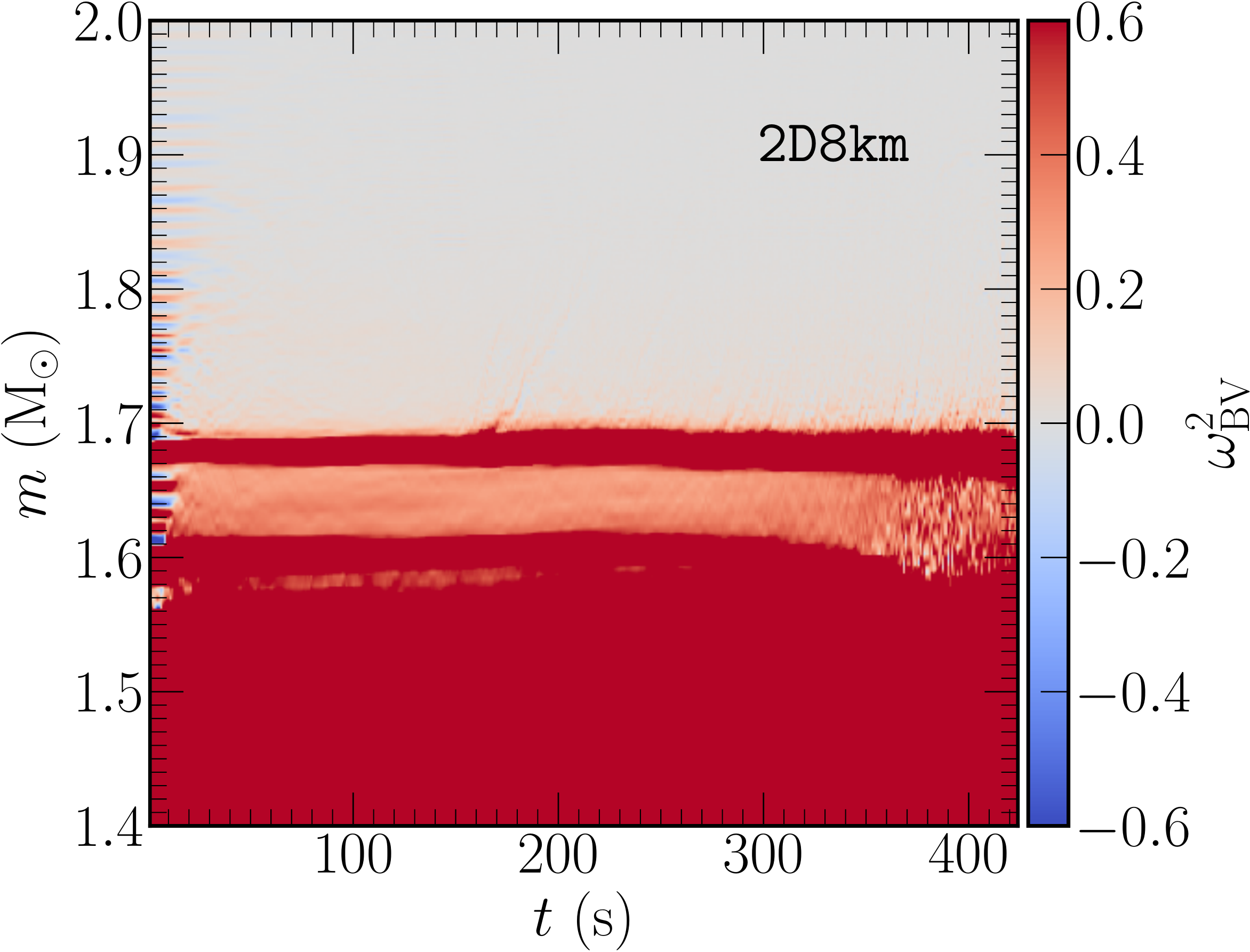}}
        \end{subfigure}
        \begin{subfigure}{
                \includegraphics[width=0.47\textwidth]{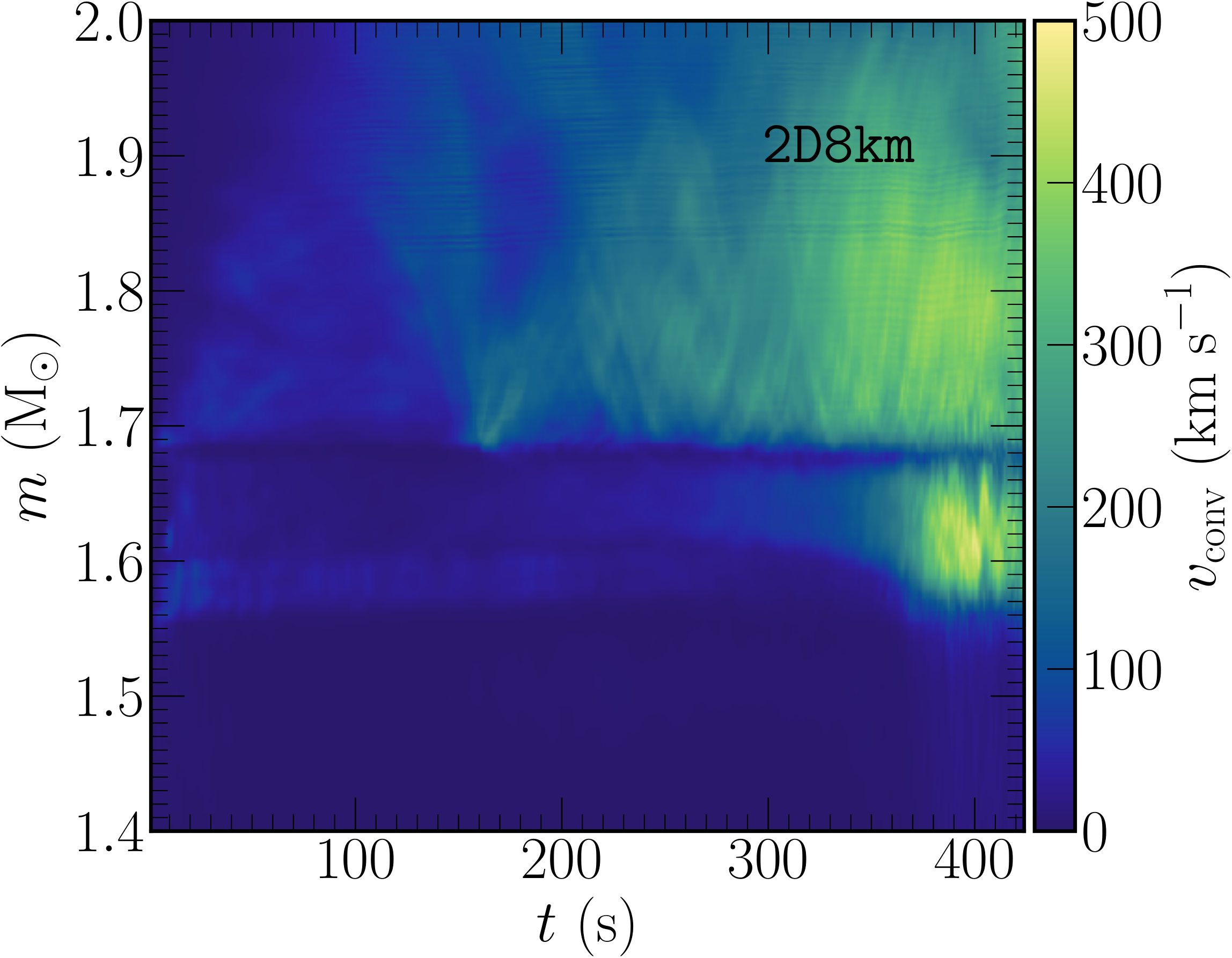}}
        \end{subfigure}
        \caption{
Same as in Figure~\ref{fig:1d_conv_time_evol} but for the \texttt{2D8km} \FLASH simulation. The 
slightly negative values of $\omega^{2}_{\rm{BV}}$ denoted by the gray/light blue regions represent 
regions unstable to convection.
        }\label{fig:2d_conv_time_evol}
\end{figure*}

In Figure~\ref{fig:2d_time_evol_max_mach_number_shells} we show the time evolution of the maximum 
angle-averaged Mach number for all 2D models. The maximum Mach number 
reported at the start of the simulation reflects the initial transient as it traverses the domain. Beyond 
$t\approx15$ s, the transient has either traversed the shell region or been sufficiently damped that the Mach 
numbers reflect the convective properties of the shells and not the initial radial wave.  
The Si-shell appears to reach a quasi-steady state in the first $\approx$ 80 seconds, this is seen by all models 
reaching a characteristic Mach number within the shell. In the O-shell, the Mach number remains relatively flat 
although larger than the approximate mean in the Si-shell, this suggests little to no convective activity in the O-shell
during this time. To make an estimate of the time at which these two regions would be expected to reach a quasi-steady
convective state, we can estimate a convective turnover timescale for each region. The Si-shell spans a radius of 
approximately 800 km with convective speeds of $v_{\rm{Si-shell}}\approx 80$ km s$^{-1}$ at early times. Using this, 
we can estimate an approximate convective turnover time within the Si- shell of $\tau_{{\rm{Si}}} \approx 2 r_{\rm{Si}} / v_{\rm{Si}} \approx 20$ seconds. This value suggests that after the transient has traversed the Si-shell region, a 
total of approximately three turnover timescales elapse before the region reaches a quasi-steady state 
represented by an average Mach number oscillating on the approximate timescale of the turnover time. 
A similar estimate can be made to the O-shell where we determine a turnover time of 
$\tau_{\rm{O}}\approx$ 100 seconds. This value suggests that the lack of change in Mach number for the O-shell
is due to the fact that the region has not yet reached a quasi-steady convective state. 

\added{
Figure \ref{fig:2d_time_evol_max_mach_number_shells} also shows that the time evolution of the maximum Mach number for the 2D simulations is not monotonic with resolution. 
While up to around 160 s the peak Mach number is consistently larger for lower resolution simulations, thereafter we find a more varied and chaotic trend. 
The maximum Mach number does not show a clear trend or behavior with resolution in the later parts of the simulations.
We attribute this to the stochastic behavior of the turbulent convection, particularly in 2D where the role of the inverse turbulent cascade \citep{kraichnan:1967} and the influence of the symmetry axis can exacerbate chaotic behavior. 
In the highest
resolution 2D model, the O-shell reaches speeds of $v_{\rm{conv.}} \approx 250$ km s$^{-1}$, slightly larger than 
the 16 km model. This is seen as a divergence of the two models at $t \approx300$ s with the more resolved model 
increasing in maximum Mach number value, reaching values that the 16 km model does not reach until the last 
$\approx 15$ seconds prior to collapse. The 
\texttt{2D24km} has the largest convective velocity speeds at $t \approx 300$ s with speeds continuing to increase 
up to the point of collapse, this trend is reflected as the steady increase of the local maximum Mach number. The lowest
resolution 2D model has convective speeds $\approx 100$ km s$^{-1}$ slower than the \texttt{2D24km} at this time 
with speeds maintaining steady values up until the final $\approx 20$ s before collapse, represented by the 
nearly constant Mach number of $\approx$ 0.07. In the final $\approx$ 20 s, the \texttt{2D32km} only slightly 
increases to a value of $\approx0.08$ at collapse. }
By the end of the 
simulation, the models span a range of Mach numbers of $\approx0.08-0.12$.

The late time behavior of the 
Si-shell region of the two highest resolution models can be attributed to an expansion of the width of the 
convective Si-shell region observed in the last slice of Figure~\ref{fig:2d_si28_time_evol} \added{as well as a steady increase
followed by saturation of the peak mach number in Figure~\ref{fig:2d_time_evol_max_mach_number_shells}}. 
\added{The two highest resolution models experience a less severe initial and reflected velocity transient wave 
($\approx 20$ km \ s larger in the 32 km model than the 8 km model), 
causing less deviation from hydrostatic equilibrium for the Si-shell region as opposed to the less resolved models.
Due to less expansion in} the two highest resolution models, the convective velocity speeds reach \added{larger
values in the Si-shell at earlier times, this is represented by a notable increase in the mach number from 
$t\approx 300$ s to $\approx 390$ s when a peak value or ``saturation'' point is reached. At this point, 
speeds in these models become large} enough to
overcome the barrier between the convective and non-convective silicon shell regions causing them to merge. 
After this merging of these two regions, the entire region becomes fully convective. However, due to the expansion 
of the width of the Si-shell region after merging, the burning within this region occurs at lower density. Because the 
local sound 
speed goes as $c_{\rm{s}}\propto \rho^{-1/2}$, a decrease in density leads to an increase in the sound speed 
therefore decreasing the local maximum Mach number. The two lower resolution models do not experience this 
merging of the two regions until moments before collapse when the flow speeds are large enough $\approx$ 150 
km s$^{-1}$ to merge. 

\begin{figure}[!htb]
\centering{\includegraphics[width=1.0\columnwidth]{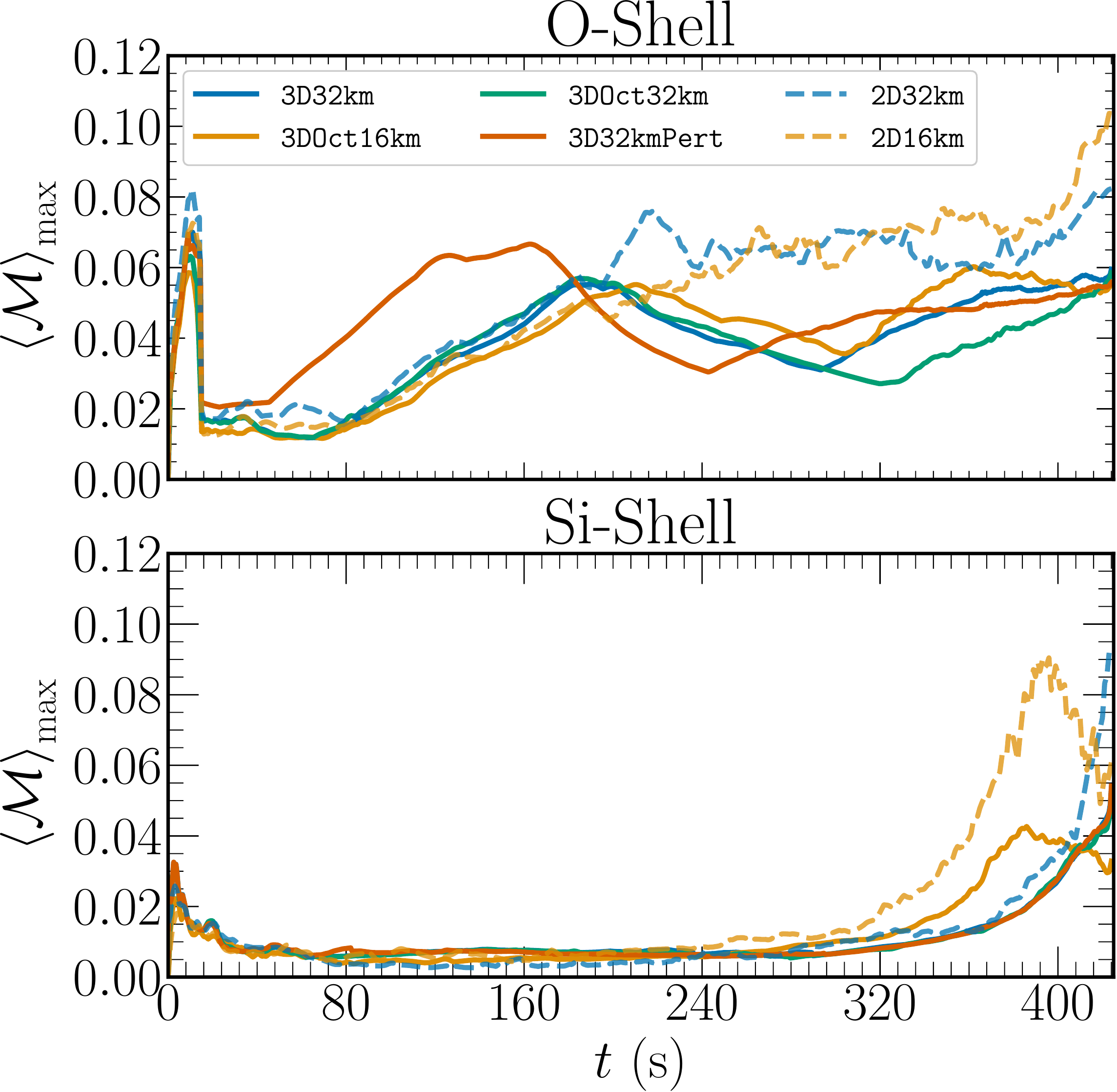}}
\caption{
Same as in Figure~\ref{fig:2d_time_evol_max_mach_number_shells} but for all the 3D models for the 
duration of the simulation. \added{For comparison, we also plot the \texttt{2D16km} and \texttt{2D32km} 
models.}
}\label{fig:3d_time_evol_max_mach_number_shells}
\end{figure}

In Figure~\ref{fig:2d_ekin_total_time_evol} we show the time evolution of the total kinetic energy in the radial
and non-radial components for the 2D \FLASH models. When considering the non-radial kinetic energy
components for the four models we see that the peak value of the energy at collapse increases with an
 increase in model resolution with the highest resolution model showing a peak value of 
$E_{\rm{kin.},\theta} \approx 2.5\times10^{47}$ erg at collapse. The radial kinetic energy shows further 
evidence for the expansion of the Si-shell region in the two highest resolution models with the energies 
showing local maxima around $t\approx$ 390 s followed by a steady decline for the duration of the 
simulation. This transition time is also reflected in the non-radial kinetic energy where one can notice a slight 
increase in the energy from $t\approx 390$ s to collapse.

\begin{figure}[!htb]
\centering{\includegraphics[width=1.0\columnwidth]{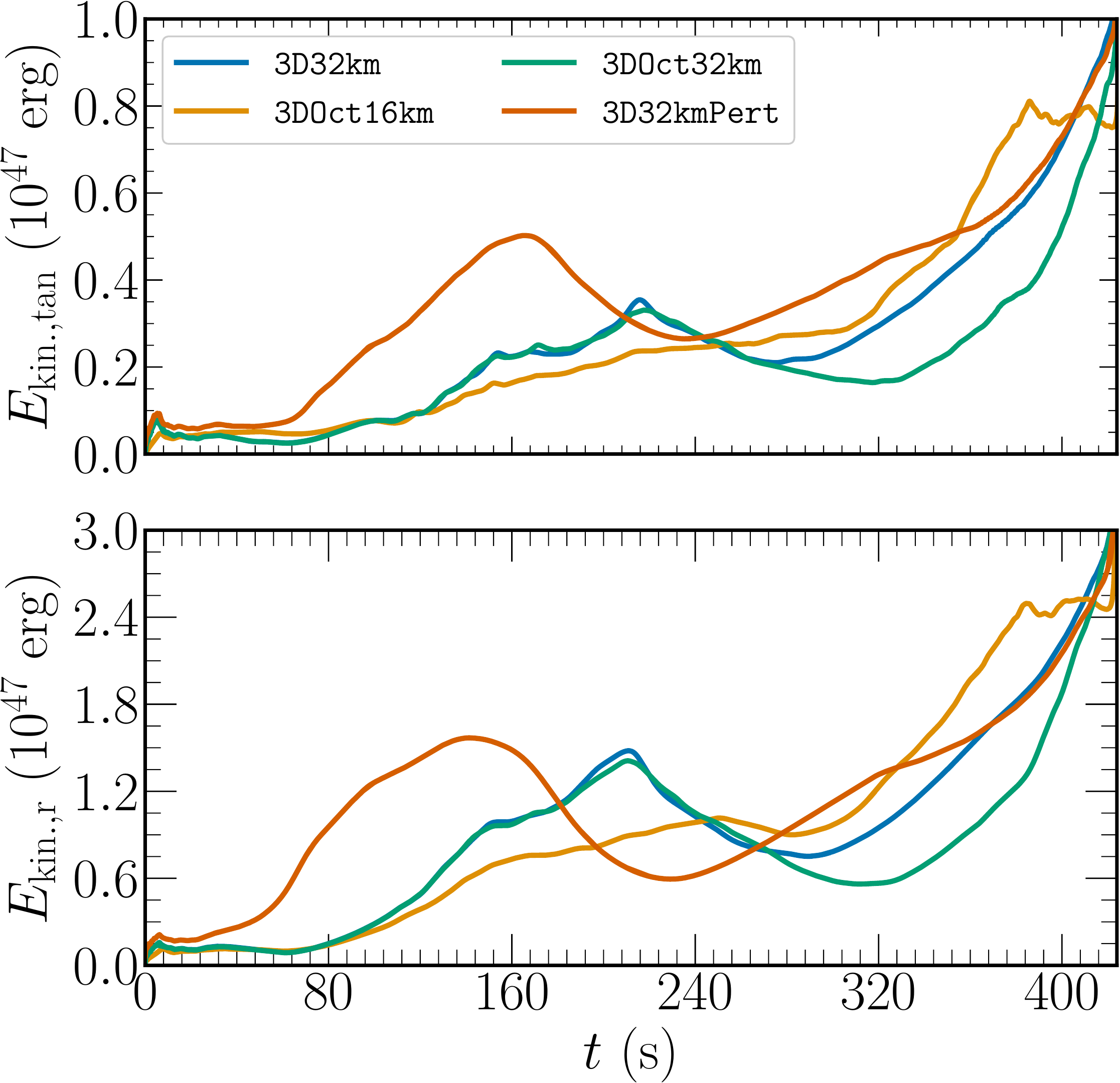}}
\caption{
Same as in Figure~\ref{fig:2d_ekin_total_time_evol} but for the four 3D models.
}\label{fig:3d_ekin_total_time_evol}
\end{figure}

In Figure~\ref{fig:2d_conv_time_evol} (left) we show the time evolution of the \BV\ frequency for the 
\added{\texttt{2D8km} model} assuming 
the Ledoux criterion for convection which states that a region is stable against convection if 

\begin{equation}
\nabla_{\rm{rad.}} < \nabla_{\rm{ad.}} - \frac{\chi_{\mu}}{\chi_{T}} \nabla_{\mu}~.
\end{equation}
Under this criterion, we can compute the \BV\ frequency for the \FLASH simulations as
\begin{equation}
\omega^{2}_{\rm{BV}} = g \left ( \frac{1}{\rho} \frac{\partial{\rho}}{\partial{r}} -  
\frac{1}{\rho c^2_{\rm{s}}}  \frac{\partial{P}}{\partial{r}}~ \right )~,
\label{eq:BV}
\end{equation}
where $g$ is the local gravitational acceleration, $\rho$ the mass density, $c_{\rm{s}}$ the adiabatic sound
speed, and $P$ the pressure. This form of the \BV\ frequency equation is equivalent to 
 forms that explicitly include the entropy and electron fraction gradients \citep{muller_2016_aa}. 
For each timestep for which these 2D data are available, we compute angle 
average profiles as a function of radius before using Equation~\ref{eq:BV} to compute
the frequency. Using this convention a positive value implies a region stable against convection. 
\added{ In multi-D simulations, the asymptotic behavior of efficient convection at evolutionary timescales will work 
to flatten or smooth out any entropy or composition gradients thus tending towards a \BV\ frequency of zero. 
In the context of this work, we are using this quantity to identify the location of convectively active regions 
and compare them amongst models and to other multi-D progenitor models found in the literature 
\citep{muller_2016_aa, jones_2016_aa,yadav_2019_aa}.}

In Figure~\ref{fig:2d_conv_time_evol} we also show the time evolution of the convective velocity (right), 
here defined as $ v_{\rm{conv}} =  \sqrt{ |\textbf{v}| ^2 - v^2_{\rm{rad.} }}$, as a function of time for the same
model. The base of the O-shell region is shown at an approximate mass coordinate of 1.7 \msun. Within the 
O-shell, the convective velocity reach speeds of nearly 500 km s$^{-1}$ as the model approaches iron core collapse. 
Prior to this, the model shows values on the order of 50-100 km s$^{-1}$ in the Si-shell and 200-400 km s$^{-1}$ in the O-shell. 
The expansion of the convective Si-shell region due to the merging is observed as well in the convective velocity, 
again around $t\approx390$ s, the same time at which the velocity begins to reach values observed in the 
O-shell for this model. In comparing to Figure~\ref{fig:1d_conv_time_evol}, the \FLASH model predicts two initial
inner and outer convectively active regions that eventually merge into one larger region near collapse. On the
other hand, the \MESA model predicts a transition of the location of the convective region followed by late time 
expansion of this region near collapse. Despite these somewhat different evolutionary pathways both models
agree in their qualitative description of the location of the convective regions near collapse. The major difference
between the two models are the magnitude of the convective velocities predicted by \MESA/MLT.

\begin{figure*}[!htb]
         \centering  
        \begin{subfigure}{
                \includegraphics[width=0.85\textwidth]{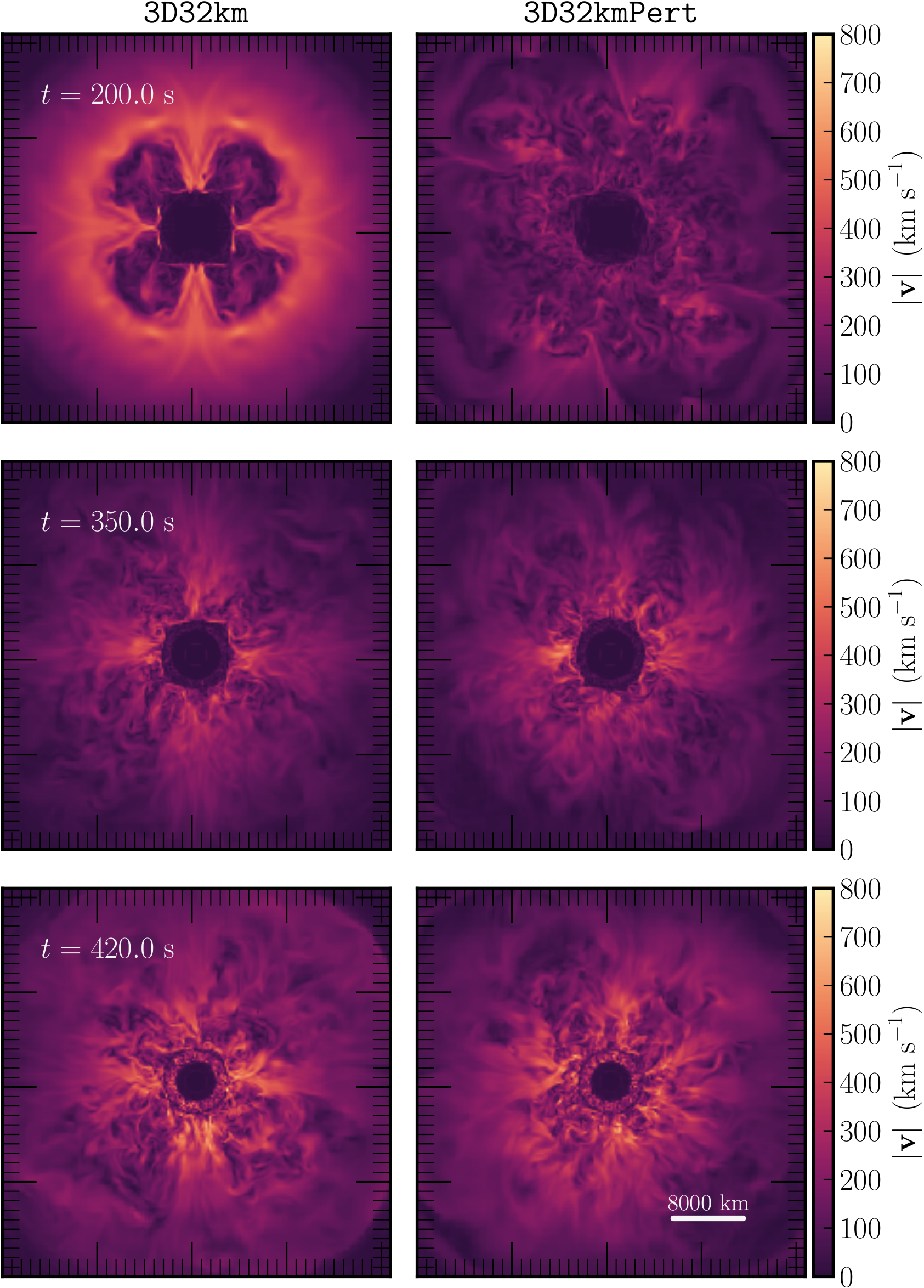}}
        \end{subfigure}
        \caption{
Slices of the the magnitude of the velocity field for the \texttt{3D32km} (left) and \texttt{3D32kmPert} (right) models 
at $t=$ 200 s, 350 s, and 420 s in the $x-y$ plane. \added{At a time of $t=$ 200 s, the unperturbed \texttt{3D32km} (left)
model shows a clear indication of grid scale perturbations seeded by the Cartesian grid while the perturbed
model represents the imposed initial velocity field described in \S~\ref{sec:methods_mD}. 
As the simulation evolves, the boxiness is lessened as the convection becomes fully developed in the \texttt{3D32km} model, 
though still remains to some extent at the end of the simulation. At $t=420$ s (bottom panel), the qualitative structure of 
the O shell convection in both models  is similar, though \texttt{3D32km} does still 
show a greater impact of the Cartesian grid. Overall, both models show O-shell convection occurring at a broad range 
of scales with the unperturbed model having a slight preference along the Cartesian axis. The characteristic scale of the 
convection for both models is considered in \S~\ref{sec:spectra}.}
        }\label{fig:3d_mag_vel_time_evol}
\end{figure*}

\subsection{Results from 3D Simulations}
\label{sec:3d}
We perform a total of \added{four} 3D stellar models: two models in octant symmetry at 16 and 32 km finest grid spacing 
and \added{two} full 4$\pi$ models at 32 km finest resolution \added{with and without initial perturbations}. In this subsection, 
we will consider the global properties of all 3D 
models, describe the perturbed 4$\pi$ 32 km in detail, and, lastly, consider the impact of octant symmetry and resolution.

In Figure~\ref{fig:3d_time_evol_max_mach_number_shells} we show the time evolution of the maximum 
Mach number in the Si- and O-shell at each timestep for all 3D models. Contrary to the trend seen in Figure~\ref{fig:2d_time_evol_max_mach_number_shells} for the O-shell one can observe a periodic nature in the 
Mach number values that follows our estimate for the convective turnover timescales from Section \ref{sec:2d}.
Similar to the comparable 2D case (see Figure~\ref{fig:2d_ekin_total_time_evol}), the \texttt{3DOct16km} model appears to reach a peak 
Mach number in the Si-shell at around $t\approx390$ s before a steady decline as the model approaches 
collapse, this behavior is also attributed to expansion of the Si-shell after the merging of the convective and
non-convective regions. This result suggests that the merger is independent of geometry and dimensionality but instead depends on the resolution of the inner core region. 

Figure~\ref{fig:3d_ekin_total_time_evol} shows the time evolution of the radial and non-radial kinetic 
energy for the \added{four} 3D models. In general, the 32 km resolution models behave similarly with the 
4$\pi$ models having larger kinetic energy values than the octant model beyond $t\approx280$ s. 
\added{The perturbed 4$\pi$ 3D model reaches larger kinetic energy values at earlier times due to the initial injection
of energy via the perturbations. The model reaches a peak in radial kinetic energy at $t\approx$ 150 s, roughly 70 s
earlier than the unperturbed model.  In terms of the total initial kinetic energy the perturbed model began with a value of 
$E_{\rm{kin.}} \approx 5.8\times10^{45}$ erg, approximately 4\% of the peak total kinetic energy observed later.}
The 16 km octant model has larger kinetic energy values 
than both models at late times while also reaching a local peak value approximately 30 s before collapse.
Prior to about 200 s, the bulk of the energy in the radial direction for the two unperturbed 32 km models is 
due to an the initial transient that
traverses the domain at the start of the simulation and leaves the domain at $t\approx $ 60 s.
The 16 km model experiences
a less significant initial transient and therefore undergoes less initial expansion/contraction as 
the model readjusts to a new equilibrium.

\begin{figure*}[!htb]
         \centering
        \begin{subfigure}{
                \includegraphics[width=\textwidth]{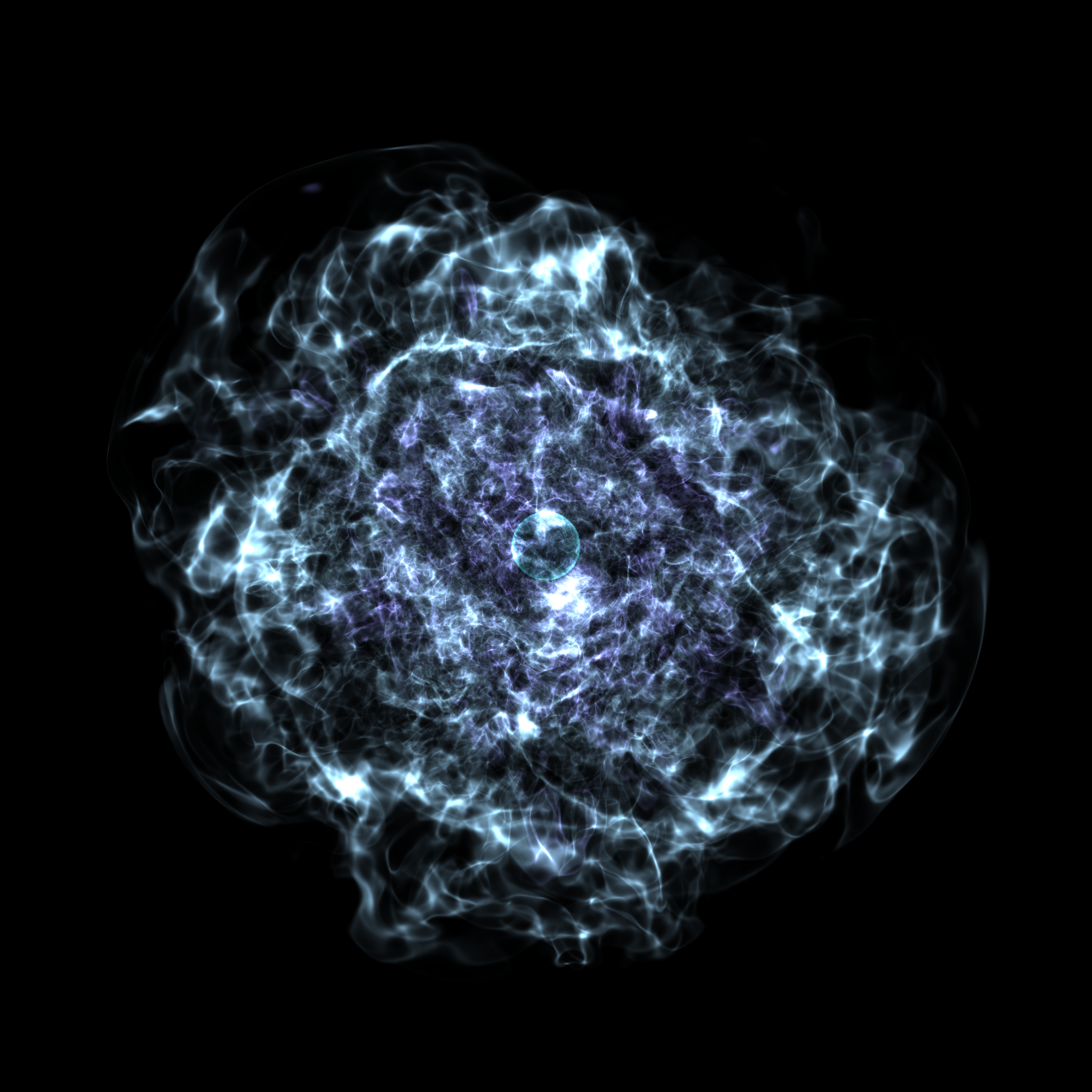}}
        \end{subfigure}    
        \caption{
3D volume rendering of the magnitude of the velocity field for the 4$\pi$ \texttt{3D32kmPert} model at a time 
$\approx$ 2 seconds before iron core collapse. The teal contour denotes the edge of the iron core defined 
to be the radius at which $s\approx$ 4 $k_{\rm{KB}}$ / baryon. At this time, the iron core spans a radius of $r\approx 1982$ km. 
The light purple plumes represent O-shell burning convection speeds following a Guassian \added{intensity mapping} 
with a peak at $\left | \bf{v} \right| \approx 300$ \added{km s$^{-1}$}  while the light blue plumes \added{use a similar mapping with a peak at}
$\left | \bf{v} \right|\approx 100$ \added{km s$^{-1}$. Both intensity mappings have a width of $\pm$ 30 km s$^{-1}$ about the 
peak of the Gaussian curve.} This image was made using \texttt{yt} and the color map 
library \texttt{cmocean}.
        }\label{fig:mag_vel_vol_rend}
\end{figure*}

The evolution of the magnitude of the velocity field for models \texttt{3D32km} (left) and \texttt{3D32kmPert} (right) is compared in 
Figure~\ref{fig:3d_mag_vel_time_evol}.  
\added{For the unperturbed case, \texttt{3D32km}, convection is seeded by the Cartesian grid, resulting in a prominent boxy character to the convection in the O shell, particularly at earlier times. 
The boxiness is lessened as the convection becomes fully developed, though still remains to some extent at then end of the simulation.
This is seen in the bottom left panel of Figure \ref{fig:3d_mag_vel_time_evol} as slightly larger velocities along the directions normal to the Cartesian grid.
This is to be expected since the effective numerical viscosity is larger in the diagonal directions, damping the convective velocities there. 
Use of high-order accurate methods, both in space and time, could alleviate this Cartesian grid effect to some extent \citep[e.g.,][]{rembiasz:2017}.
For the perturbed model, \texttt{3D32kmPert}, vigorous convection begins earlier, as evidenced in Figure \ref{fig:3d_ekin_total_time_evol}, and does not show clear Cartesian grid artifacts, as can be seen in the top right panel of Figure \ref{fig:3d_mag_vel_time_evol}.
As the simulation proceeds, the influence of the Cartesian grid is still seen in \texttt{3D32kmPert} to some extent resulting in slightly larger velocities along the normal directions, though still to a lesser degree than in \texttt{3D32km}.
Just prior to collapse (bottom panels of Figure \ref{fig:3d_mag_vel_time_evol}), the qualitative structure of the O-shell convection in models \texttt{3D32kmPert} and \texttt{3D32km} is not too dissimilar, though \texttt{3D32km} does still show a greater impact of the Cartesian grid. 
The total energy in convection is also nearly identical between the two models near collapse, as seen in Figure \ref{fig:3d_ekin_total_time_evol}. 
The Si shell convection shows little to no Cartesian grid effect and is essentially identical between models \texttt{3D32kmPert} and \texttt{3D32km}. In Section \ref{sec:spectra} we consider properties of the convection in these two regions more
quantitatively by decomposing the velocity field into spherical harmonics and comparing those results between the 
perturbed and non-perturbed models. 
}

In Figure~\ref{fig:mag_vel_vol_rend} we show a 3D volume rendering of the magnitude
of the velocity field for the \texttt{3D32kmPert} model at $t\approx 423$ s. The approximate location of the 
edge of the iron core (shown in teal) is taken to be an isocontour surface at a radius where 
$s\approx$ 4 $k_{\rm{KB}}$ / baryon. At this time, the 
iron core has a radius of $r\approx 1982$ km. The light purple plumes show the fast moving convective 
motions in the O-shell region depicted by a Guassian transfer function with a peak at 
$\left | \bf{v} \right| \approx 300$ \added{km s$^{-1}$}. The slower moving, larger scale motions are shown using a similar 
transfer function with a peak at $\left | \bf{v} \right|\approx 100$ \added{km s$^{-1}$} in light blue. \added{Both intensity 
mappings have a width of $\pm$ 30 km s$^{-1}$ about the peak of their respective Gaussian curves. The volume
rendering suggests that convection is occurring in the O-shell at a broad range of convective scales.}

\begin{figure*}[!htb]
         \centering  
        \begin{subfigure}{
                \includegraphics[width=0.482\textwidth]{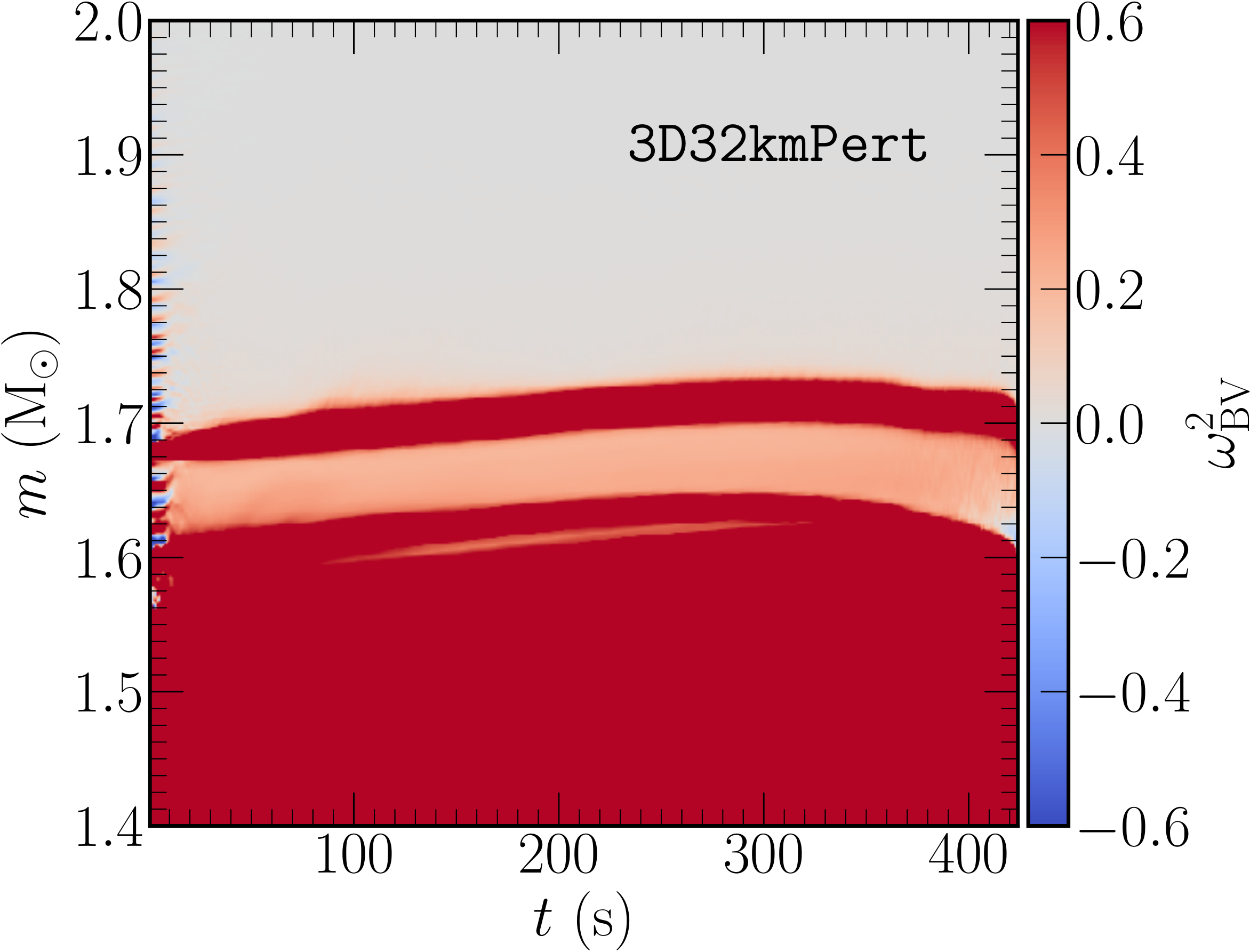}}
        \end{subfigure}
        \begin{subfigure}{
                \includegraphics[width=0.47\textwidth]{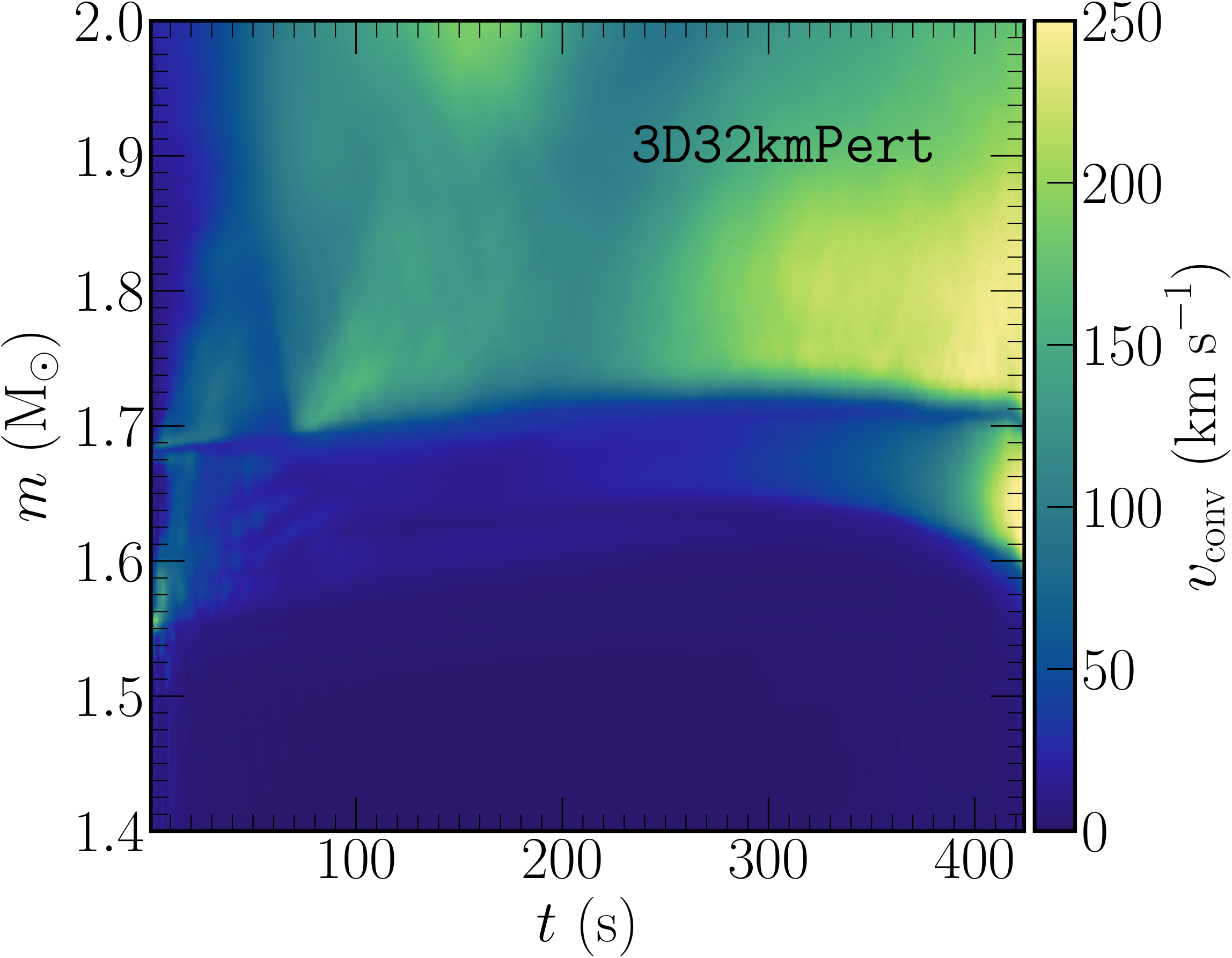}}
        \end{subfigure}
        \caption{
Same as in Figure~\ref{fig:1d_conv_time_evol} but for the \texttt{3D32kmPert} \FLASH simulation. 
        }\label{fig:3d_conv_time_evol}
\end{figure*}

Similar to the analysis done for the \texttt{MESA} model and the 2D \FLASH models, in Figure~\ref{fig:3d_conv_time_evol} 
we show the \BV\ frequency (left) and convective velocity (right) as a function of time for the \texttt{3D32kmPert} \FLASH model.
Unlike the \texttt{2D8km}, this model does not experience the merging of the two Si regions until a few seconds prior to 
collapse leading to a similar fully convective Si-shell at the end of the simulation.
Another notable feature of this model is the slight expansion and then contraction of different 
regions of the model. For example, the base of the O-shell region begins at a mass coordinate of $m \approx 1.68$ \msun
in all models but appears to expand outward to a coordinate of $m \approx 1.72$ \msun for the \texttt{3D32km} model at 
about $t \approx 200$ s. This expansion is not observed in the \texttt{2D8km} model but is partially due to the initial transient
at the beginning of the simulation, \added{similar to the low resolution 2D models in \S~\ref{sec:2d}}. The impact of this effect on 
our main results will be considered in Section~\ref{sec:resol}.

\subsubsection{Characterizing the convection in the \texttt{3d32kmPert} model}
\label{sec:spectra}

\begin{figure*}[!htb]
         \centering  
        \begin{subfigure}{
                \includegraphics[width=0.47\textwidth]{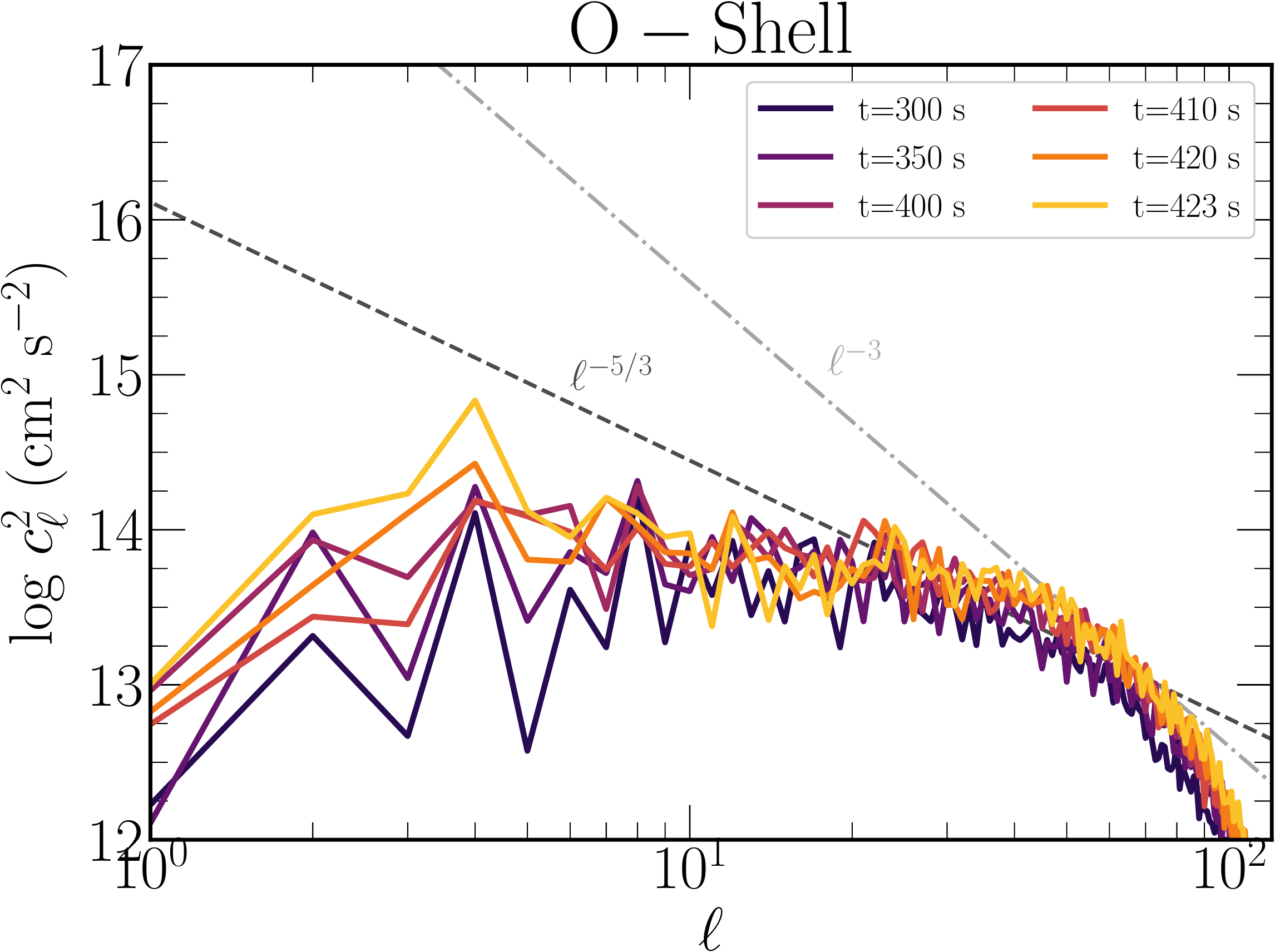}}
        \end{subfigure}
        \begin{subfigure}{
                \includegraphics[width=0.47\textwidth]{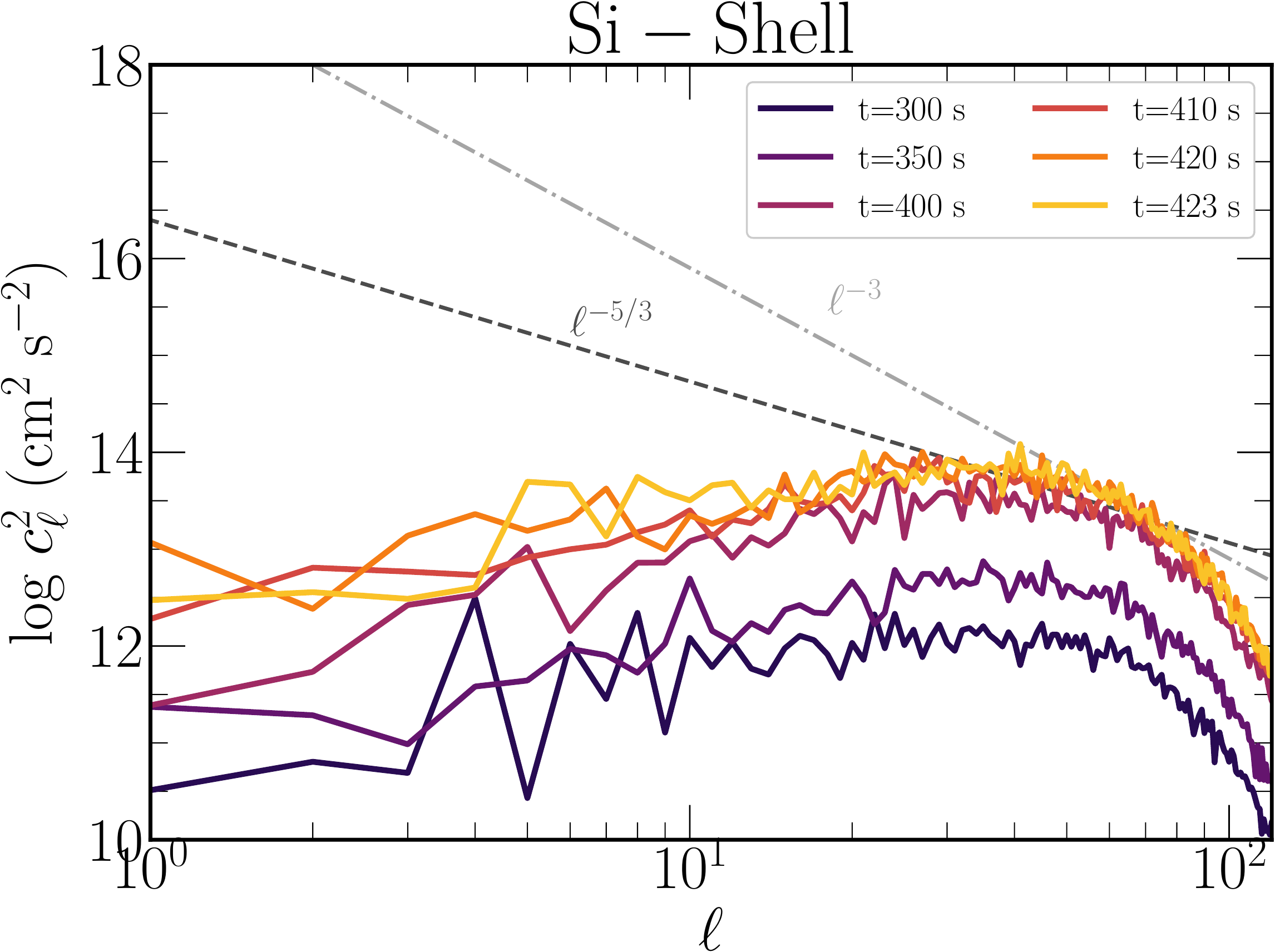}}
        \end{subfigure}
        \caption{
\added{Power spectrum distribution of the spherical harmonic decomposition of the magnitude of the velocity in 
the O- and Si-shell regions at six different times for the \texttt{3D32kmPert} model.}
        }\label{fig:3d_ell_spectra_time_evol}
\end{figure*}

\begin{figure}[!htb]
\centering{\includegraphics[width=0.99\columnwidth]{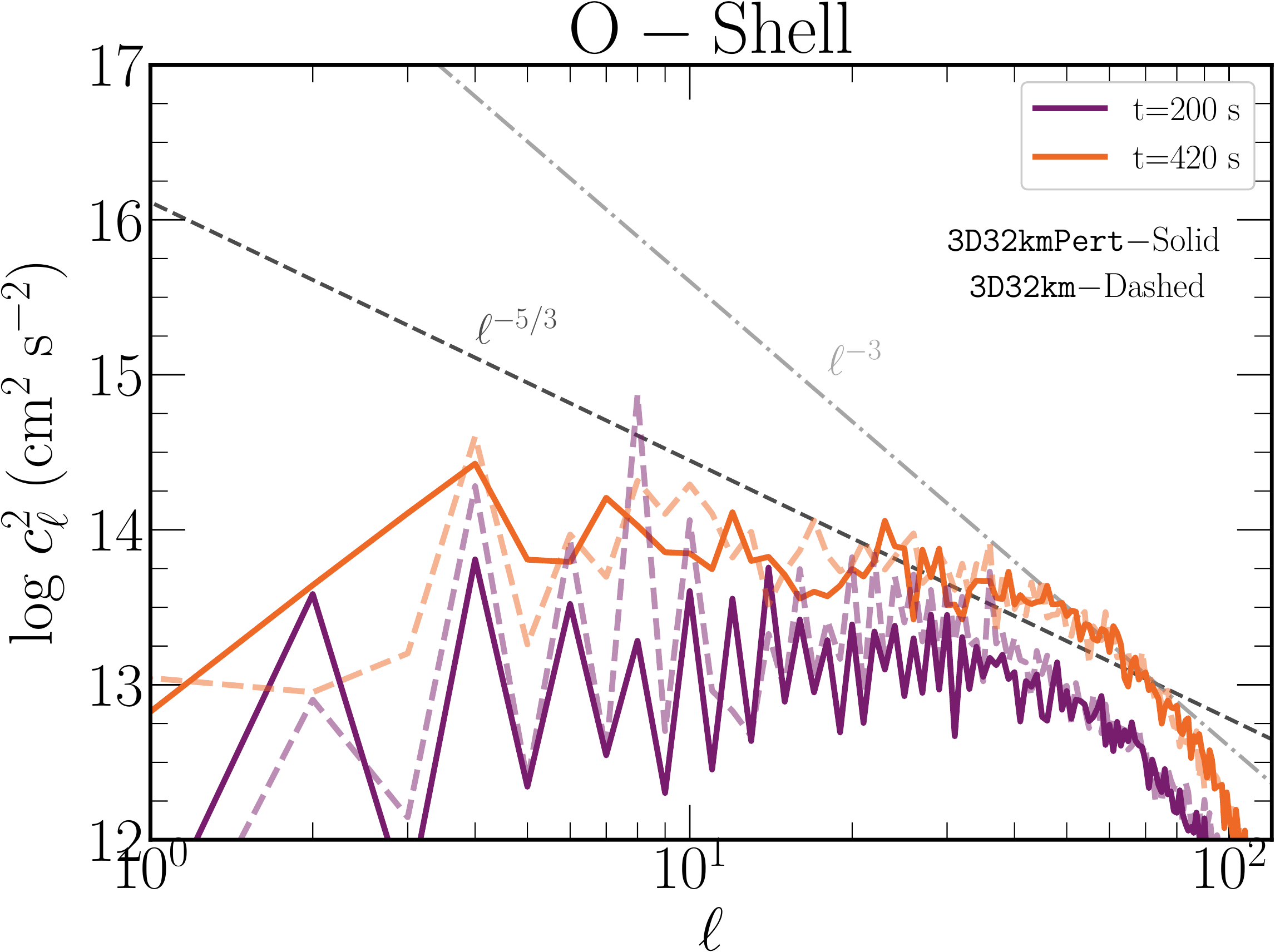}}
\caption{
Same as in Figure~\ref{fig:3d_ell_spectra_time_evol} but only considering the O-shell region for the \texttt{3D32km} 
and \texttt{3D32kmPert} models at two different times. 
}\label{fig:3d32km_vmag_o_shell_spectra_ell_compare}
\end{figure}

\begin{figure}[!htb]
\centering{\includegraphics[width=0.99\columnwidth]{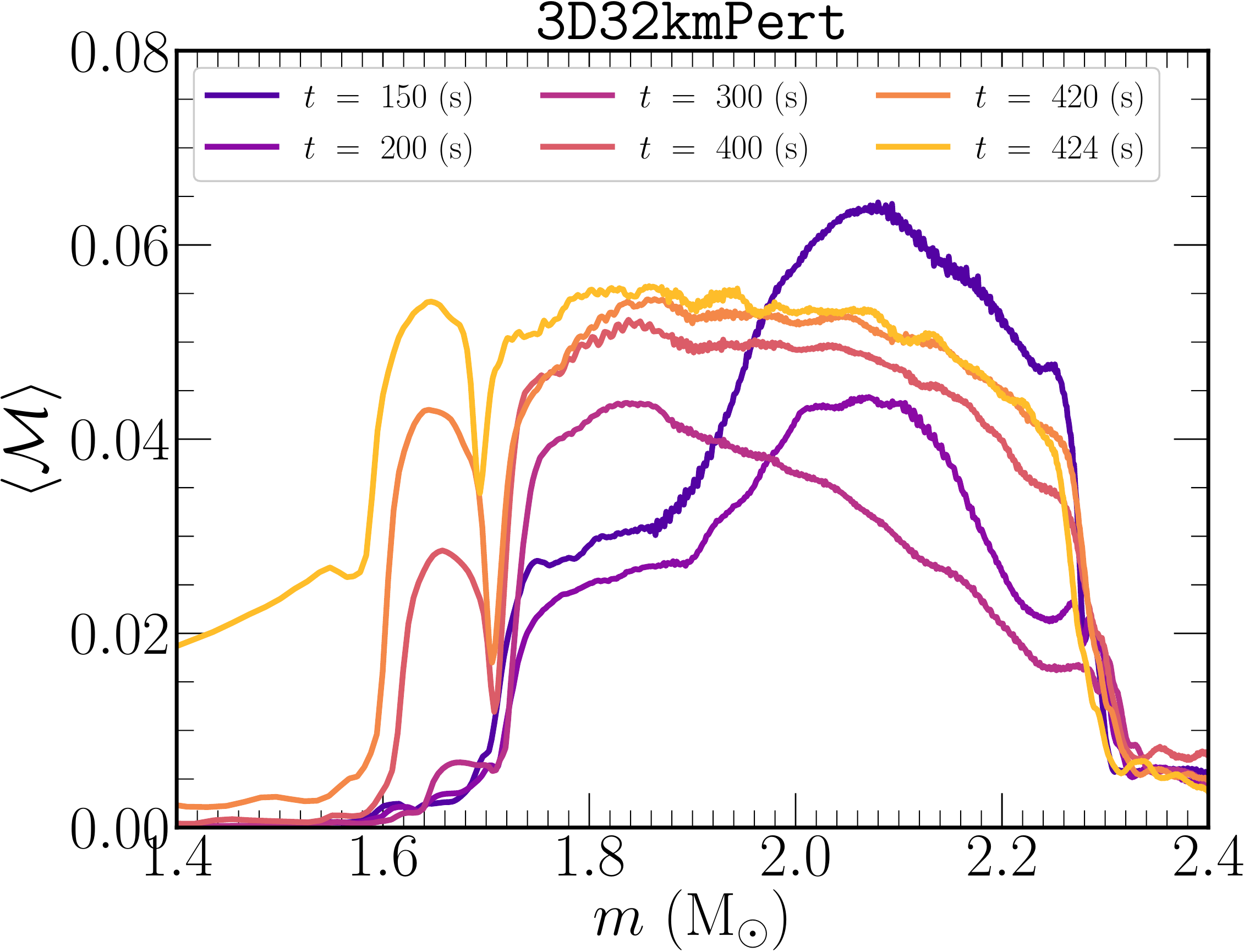}}
\caption{
The profile of the angle averaged Mach number as a function of mass coordinate at six different times for the
\texttt{3D32kmPert} stellar model. At $t=424$ s the average Mach number reaches $\left < \mathcal{M} \right > \approx 0.06$ in the
Si-shell and at the base of the O-shell.
}\label{fig:3d32km_time_evol_mach_number_mass}
\end{figure}

To characterize the scales of the convective eddies and the overall evolution of the strength of convection throughout 
the duration of the simulations we \added{decompose the perturbed velocity field into spherical harmonics for both the O- and 
Si-shells. We compute the total power for a given spherical harmonic order, $\ell$, as }
\begin{equation}
c^{2}_{\ell} = \sum\limits^{\ell}_{m=-\ell} \left | \int Y^{m}_{\ell} (\theta,\phi) 
\left | \textbf{v} \right |^{\prime}(r_{\rm{Shell}},\theta,\phi) d\Omega \right | ^{2}~,
\end{equation}
\added{where $\left | \textbf{v} \right |^{\prime}$ = $\left | \textbf{v} \right |$ - $\tilde{\left | \textbf{v} \right |}$, with $\tilde{\left | \textbf{v} \right |}$ corresponding to the mean background flow speed 
at the chosen shell radius \citep{sht_2013_aa}. For the O-shell we choose to evaluate at a radius of 6000 km, 
while the Si-shell is evaluated at $r=$ 2800 km. }

\added{ In Figure~\ref{fig:3d_ell_spectra_time_evol} we show the resulting power spectrum 
for the O- (left) and Si-shell (right) regions at six different times during the simulation for the \texttt{3D32kmPert} model.
The axis-aligned symmetry seen in the \texttt{3D32km} model at $t=200$ s of Figure~\ref{fig:3d_mag_vel_time_evol} 
is also reflected in the spectrum for the O-shell at a time of $t=300$ s here, where we see a peak at $\ell=4$ and $\ell=8$ 
indicating significant power residing in scales aligned with the Cartesian grid. 
At later times, the energy begins to transfer between scales with a noticeable increase in power in the $\ell=2$ and 
$\ell=3$ modes, while the peak of power remains at $\ell=4$. In this O-shell, we also see comparable power at 
intermediate to small scales of $\ell\approx10-30$ suggesting a range of behavior for the convection.
In the last $\approx$ 25 s, intermediate scales of $\ell=5-10$ experience an increase in 
power up towards the expected $\ell^{-5/3}$ Kolmogorov scaling (dashed black line) for turbulent flows.
The power spectrum for the Si-shell region follows a significantly different trend than in the O-shell region.  
Overall, the Si-shell region reaches a quasi-steady state at $t\approx300$ represented by a broad spectrum
with a peak at $\ell\approx30-40$. Beyond these times, the Si-shell increases in total power by $\approx$ two orders of 
magnitude with the peak of the spectrum remaining constant up to the point of collapse.}

\added{
This characteristic $\ell$ for the Si-shell approximately corresponds to the aspect ratio, 
$ \ell \approx \pi r_\mathrm{shell} / \Delta r_\mathrm{shell} = 15$, and we associate this scale with roughly the driving scale of the turbulent convection.
There is only weak evidence for a significant inertial range with Kolmogorov scaling of $\ell^{-5/3}$ in Si-shell beginning around $\ell\approx30$.
This is a far larger wavenumber for the start of the turbulent cascade than we see for the O-shell, which is to be expected since the Si-shell is much thinner. 
Similar to behavior of the spectra for the O-shell, between the driving scale and the start of Kolmogorov scaling, the spectra show behavior \added{that is consistent with the} ``bottleneck'' effect resulting from an inefficient cascade of 
turbulent energy due to insufficient resolution \citep{radice:2015}. 
At $t\geq350$ s, the power in the peak harmonic index begins to
increase significantly with smaller values of $\ell$ also experiencing an increase in power. In the last
few seconds of the simulation, an increase in power for the $\ell=5$ mode is observed although the peak of the
spectrum remains at larger $\ell$.
At these same times, the contraction of the inner iron core is accelerating leading to an increase in the Si burning rate and, subsequently, more vigorous Si-shell convection.
A similar increase in power at low $\ell$ is also seen in the O-shell during this same period. 
Evidence to support this behavior is shown by the increase in Mach number within the Si-shell observed 
at late times in Figure~\ref{fig:3d_time_evol_max_mach_number_shells} as well as the radial kinetic energy shown in Figure~\ref{fig:3d_ekin_total_time_evol}.
 }

 \added{ 
In Figure~\ref{fig:3d32km_vmag_o_shell_spectra_ell_compare}, we compare 
the power spectrum for the perturbed and non-pertubed models at $t=200$ s and $t=420$ s. At $t=200$ s, the unperturbed model (purple dashed line)
shows an excess in power at $\ell=4$ and $\ell=8$, also reflected by the Cartesian aligned nature of the 
convection shown in the top panel of Figure~\ref{fig:3d_mag_vel_time_evol}. The perturbed model (solid purple line) does
not show excess power in these modes but instead shows a range of power across modes including those
at larger scales at $\ell \leq 4$. When considering $t=420$ s, the spectrum of the O-shell region for the 
unperturbed model (dashed orange line) shows a slightly larger peak at $\ell=4$ and a deficit of power 
by an order of magnitude for the $\ell=2$ and $\ell=3$ modes. Despite these differences, the spectra 
at this time are relatively similar, both having a peak at $\ell=4$ and an intermediate range of scales 
just below what is expected for a Kolmogorov scaling of $\ell^{-5/3}$.
}

\begin{figure}[!htb]
         \centering  
        \begin{subfigure}{
                \includegraphics[width=0.47\textwidth]{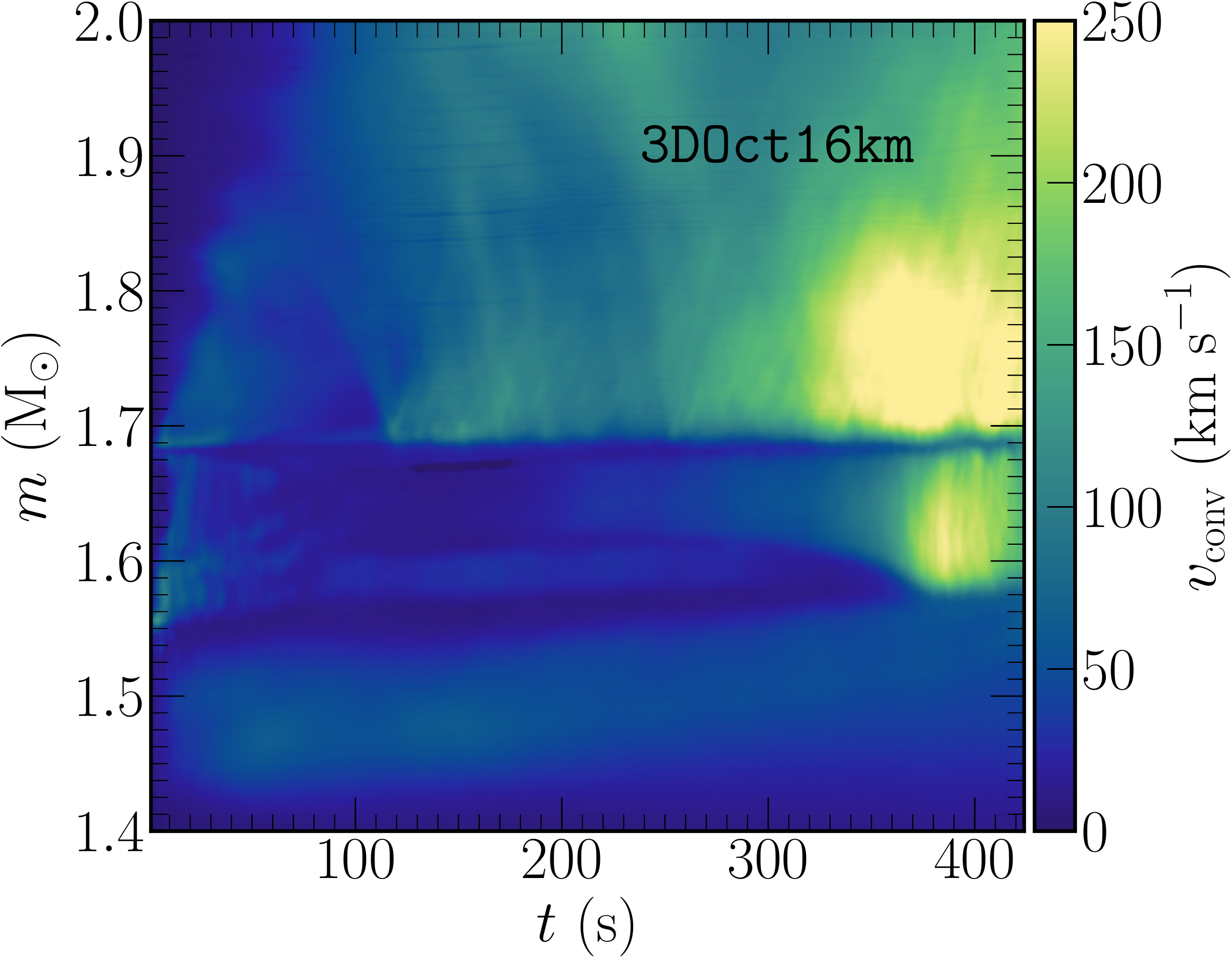}}
        \end{subfigure}
        \begin{subfigure}{
                \includegraphics[width=0.47\textwidth]{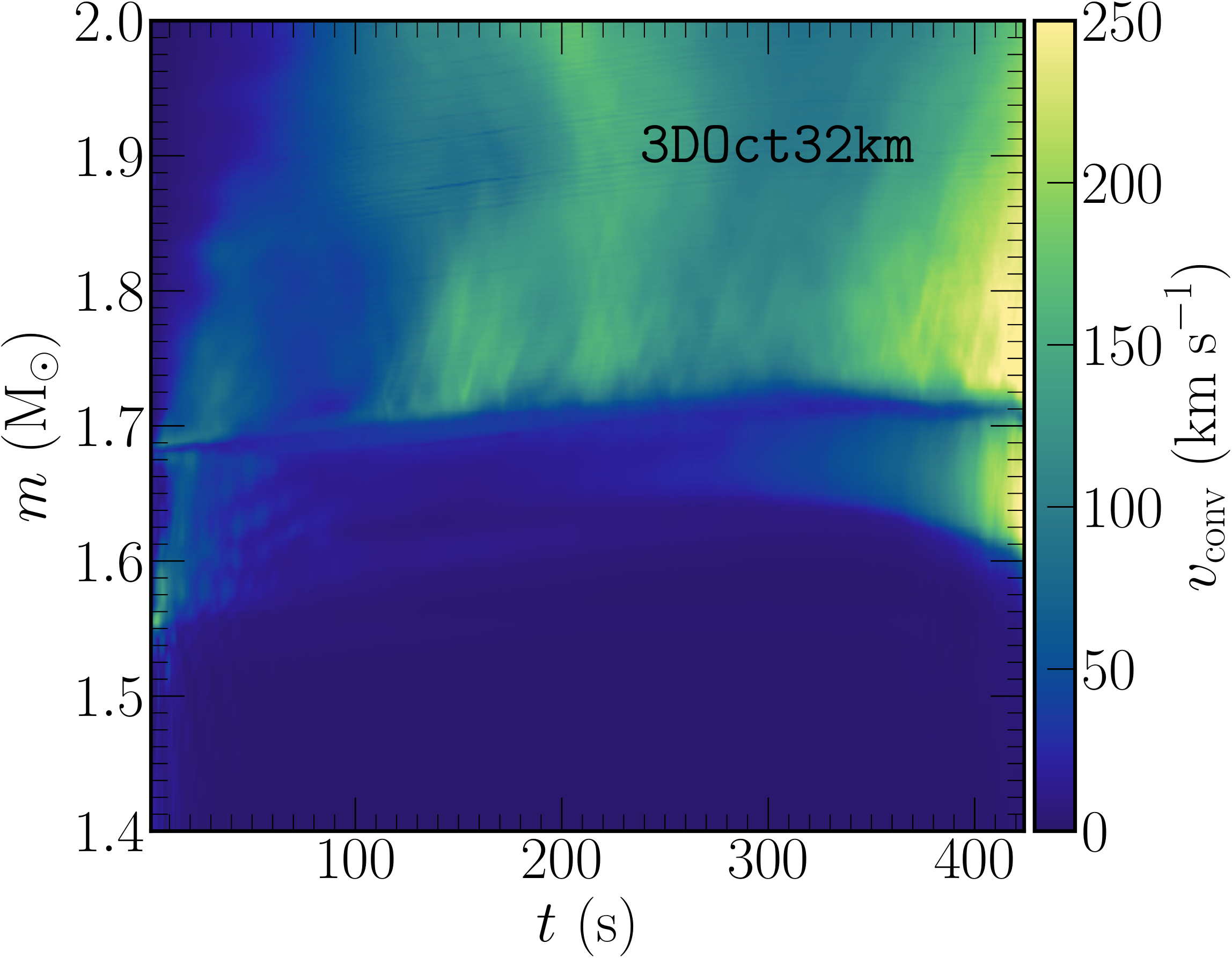}}
        \end{subfigure}
        \caption{
        The evolution of the convection velocity profiles for the 16km and 32km 3D octant models. 
        }\label{fig:3doct_v_conv_time_evol}
\end{figure}

In Figure~\ref{fig:3d32km_time_evol_mach_number_mass} we show the angle-averaged Mach number 
profile as a function of mass coordinate for the \texttt{3D32kmPert} model at six \added{different}
times. The Si-shell region is situated at 
a mass coordinate of approximately 1.6 to 1.7 \msun. The evolution of the Mach number in this region is 
further representative of the power spectra shown in Figure~\ref{fig:3d_ell_spectra_time_evol}.  
For the majority of the simulation, the Mach numbers in this region are on the order of 
$\left < \mathcal{M} \right > \lesssim 0.01$. Only at times beyond $t\approx300$ s do they increase 
significantly reaching values of $\left < \mathcal{M} \right > \approx 0.06$ prior to collapse. 
In the O-shell region, at mass coordinates of $m \approx 1.7 - 2.3\ M_{\odot}$, the Mach numbers reach values of 
about $\left < \mathcal{M} \right > \approx 0.06$ as early as $t\approx150$. At late times, 
the O-shell region approaches Mach numbers of $\left < \mathcal{M} \right > \approx 0.06$ near collapse.

\begin{figure}[!htb]
\centering{\includegraphics[width=0.99\columnwidth]{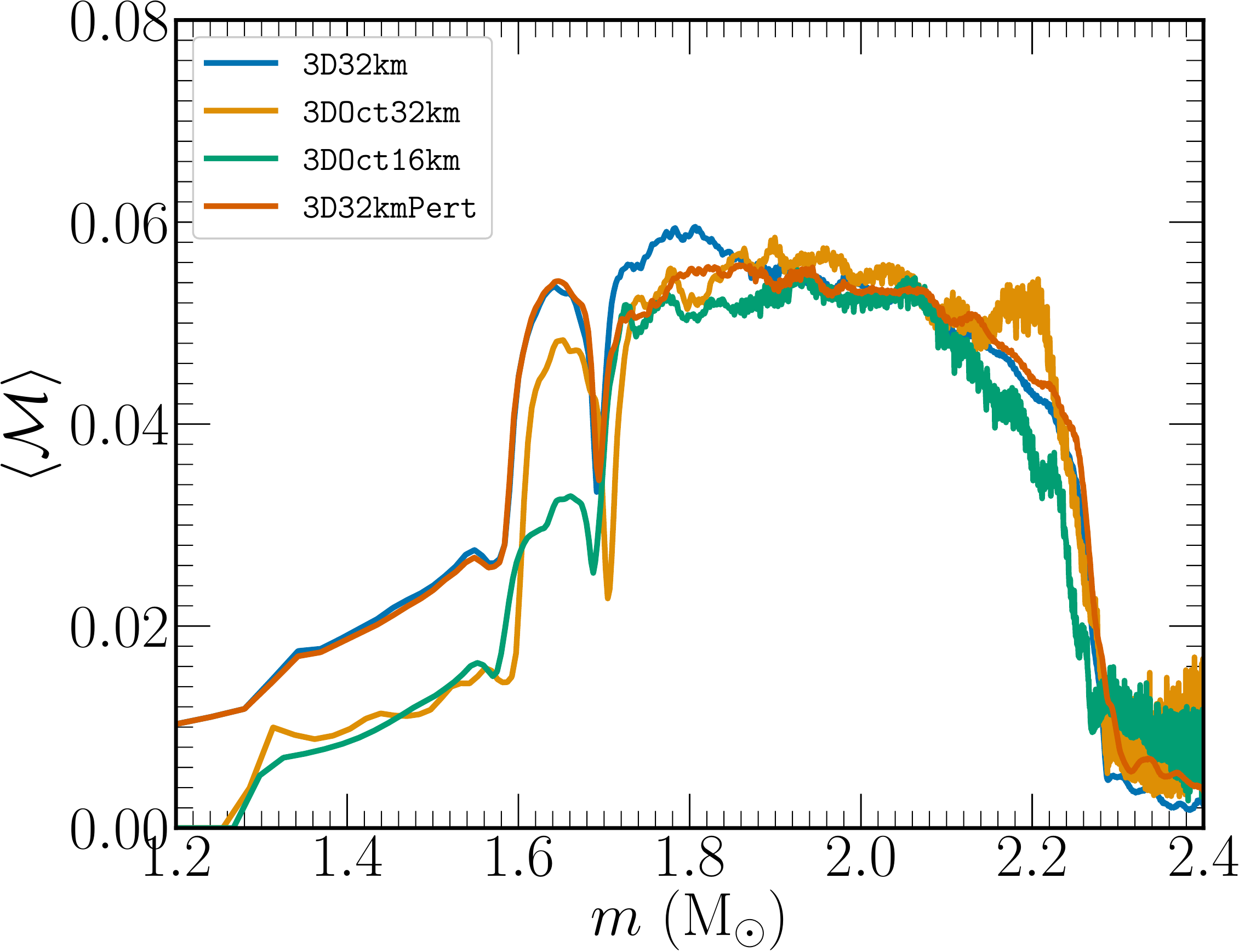}}
\caption{
The Mach number profiles as a function of mass coordinate for the four 3D models \added{at $t\approx424$ s,
the start of collapse}. 
}
\label{fig:mach_number_all_3d_models}
\end{figure}

\subsubsection{Effect of Spatial Resolution and Octant Symmetry}
\label{sec:resol}

To assess the impact of spatial resolution and octant symmetry we compare the results
of the two $4\pi$ \texttt{3D32km} simulations with the two 3D octant models. In Figure~\ref{fig:3doct_v_conv_time_evol} we 
show the time evolution of the convective velocity profiles for the \texttt{3DOct16km} and \texttt{3DOct32km} 
models. We can observe the same expansion of the O-shell region in the 32 km octant model as with the 
full 4$\pi$ models. In contrast, the 16 km octant model does not appear to undergo this contraction and the base of 
the O-shell stays at a steady mass coordinate for the duration of the simulation. Moreover, the 
16 km model reaches larger convective velocities in the Si- and O-shell regions at $t\approx$ 350 s. This 
time corresponds to the same time at which we observe a peak in the Mach numbers in 
Figure~\ref{fig:3d_time_evol_max_mach_number_shells} and is due to the merging of the convective and
non-convective regions. These results suggest that due to the stability of the O-shell in the 16 km model, 
the model follows a slightly different evolution than that of the 32 km 
octant and 4$\pi$ models characterized by larger convective velocity speeds that facilitate merging of convective
and non-convective regions in the Si-shell. Moreover, it suggests that these differences are attributed more to
the finest grid spacing of the inner core region and less dependent on the symmetry imposed for the 
octant models. Despite the differences found in the evolution of the Si-shell region between these two models,
the O-shell region appears less impacted by the difference in resolution and arrive at similar qualitative 
properties among the \added{four} 3D models. 

\begin{figure}[!htb]
\centering{\includegraphics[width=1.0\columnwidth]{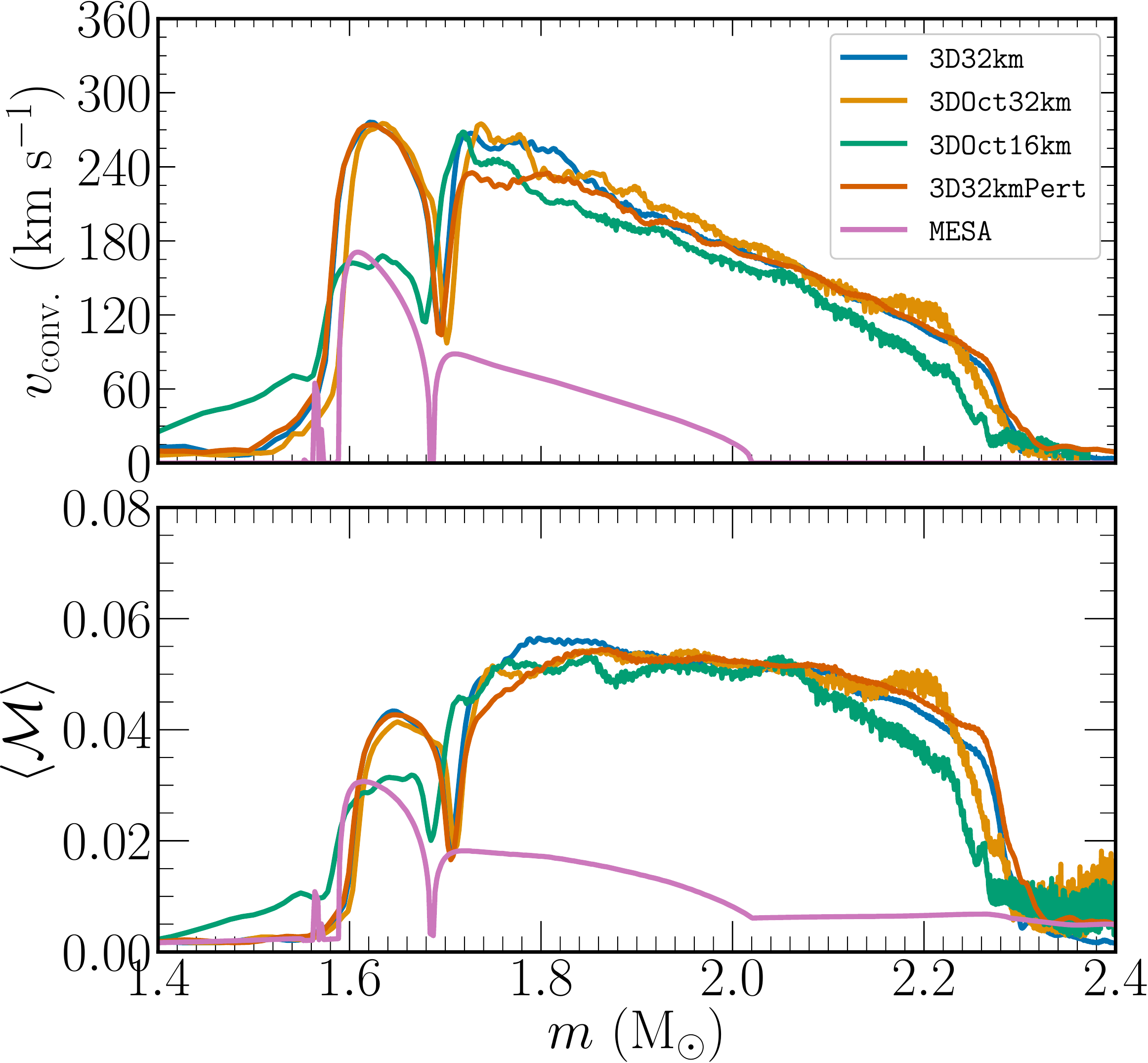}}
\caption{
The convective velocity (top) and Mach number (bottom) for the 1D \texttt{MESA} model 
and angle-averaged profiles of the four 3D models at the $t\approx 420$ s.
}\label{fig:conv_props_1_3d_comparison}
\end{figure}

We can further determine the effect that resolution and symmetry has on our results by considering some 
keys aspects of our 3D stellar models at collapse that have significant implications for simulations of CCSNe.
\citet{couch_2013_aa} considered the effect of ashpericities of imposed perturbations in the 
\added{Si}-shell regions characterized by the magnitude of the Mach number. They found large Mach number perturbations
can result in enhanced strength of turbulent convection in the CCSN mechanism, aiding explosion. 
In Figure~\ref{fig:mach_number_all_3d_models}, we plot the profiles of the Mach number at the start of collapse,
\added{$t\approx424$ s}, 
for the \added{four} 3D models.
In general, we see that the estimates of the Mach number in the O-shell region between $\approx$1.7-2.3 \msun
\added{are consistent to within $\approx 5-10\%$} across all models, 
the \added{largest difference observed} at the base of the 
O-shell in the $4\pi$ \added{unperturbed} model, showing an $\approx 8 \%$ larger value. 
The main difference is observed in the Si-shell region where the Mach number is approximately
\added{a factor of two} larger in the
32 km models, which agree with each other to $\approx$ 10\%. 

\begin{figure}[!htb]
\centering{\includegraphics[width=1.0\columnwidth]{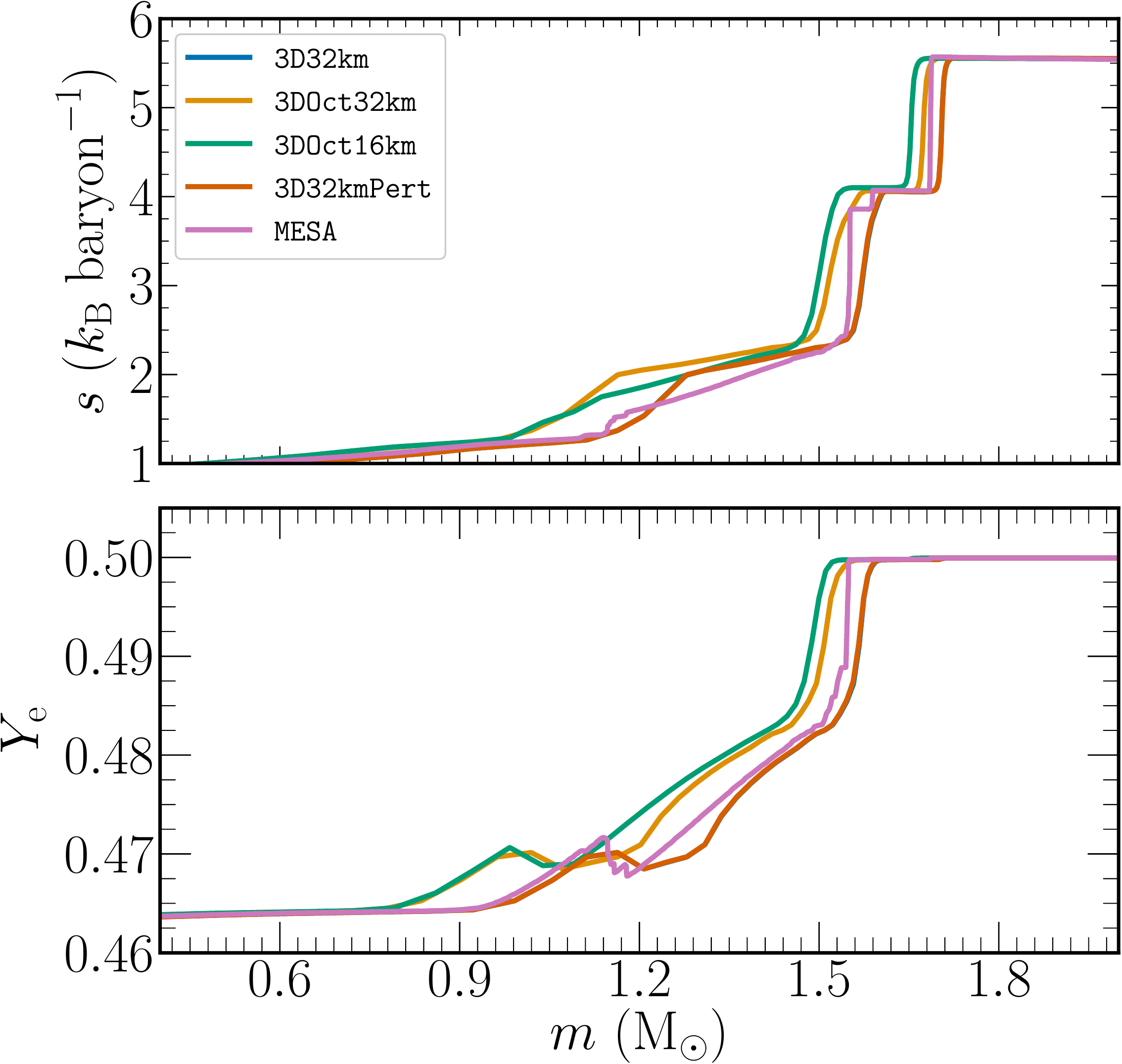}}
\caption{
The specific entropy (top) and specific electron fraction (bottom) for the 1D \texttt{MESA} model 
and angle-averaged profiles of the four 3D models at the $t\approx 420$ s.
}\label{fig:core_props_1_3d_comparison}
\end{figure}

\added{Similar to the behavior seen in the two highest resolution 2D models in \S~\ref{sec:2d},
these results suggest that the expansion of the 3D models at early times due to an initial transient can affect the subsequent evolutionary properties of the Si-shell. However, the results of the O-shell are fairly robust between resolution and symmetry. 
The $\texttt{3DOct16km}$ experiences a less severe initial radial transient leading to less expansion of the respective shell
regions allowing the Si-shell to reach higher convective speeds at earlier times than the 32 km octant model, this is 
observed Figure~\ref{fig:3doct_v_conv_time_evol}. The Si-shell then reaches a ``saturation'' point where the shell 
region expands to reach a new convective stable region with a slightly larger shell that provides for slower speeds 
at collapse and a subsequently smaller Mach number.}
When comparing the 32 km 
octant and $4\pi$ models it is likely the case that the convective speeds are able to reach larger values as 
large scale flow is not suppressed at the symmetry planes. This is supported further by the larger non-radial kinetic
energy in the $4\pi$ models as seen in Figure~\ref{fig:3d_ekin_total_time_evol}.

Another important diagnostic of the presupernova structure is the 
compactness parameter \citep{oconnor_2011_aa}, 
\begin{equation}
\xi_{m} = \left.\frac{m / M_{\odot}}{R(M_{\rm{bary}}=m) / 1000 \textup{ km}} \right\vert_{t=t_{\rm{cc}}},
\label{eqn:xi}
\end{equation}
where a value of 2.5 \msun is typically chosen for evaluation at the start of core collapse. The value of this
quantity in the progenitor star has been shown to be highly non-monotonic with ZAMS mass but gives some insight to the ensuing dynamics of the CCSN mechanism for a given progenitor \citep{sukhbold_2018_aa, couch_2019_aa}. We can compute this quantity for our three 3D models to determine
how much variation exists due to the effects of resolution and symmetry. Using Equation (\ref{eqn:xi}) for the three
3D models at a time $t=424$ s, moments before collapse. We find values of $\xi_{m=2.5M_{\odot}}=$ 0.0473, 0.0474,
0.0331, and 0.0359, for the \texttt{3D32kmPert}, \texttt{3D32km}, \texttt{3DOct16km}, and \texttt{3DOct32km} 
models respectively. These values suggest that the imposed octant symmetry 
can under estimate the compactness of the stellar model at collapse by $\approx 25\%$ while the differences 
in grid resolution but assuming octant symmetry can result in a difference of less than $\approx 10\%$. 
The compactness value at approximately the equivalent time for the
\MESA model was $\xi_{m=2.5M_{\odot}}= 0.0492$ agreeing with the $\texttt{3D32km}$ to within less than 4\%.
Our values are approximately a factor of two less than those found in \citet{sukhbold_2018_aa} and a 
factor of four less than those in \citet{sukhbold_2014_aa} \added{for a  15 $M_{\rm{ZAMS}}$ solar 
metallicity progenitor}. \added{The 1D stellar evolution model presented in this work is different from those presented in \citet{sukhbold_2018_aa} in many ways and subject to uncertainties such as input physics, mass loss rates, nuclear reactions, 
and resolution. Moreover, the complex interplay between various burning shells in the advanced burning stages
may be chaotic in 1D leading to a stochastic behavior of measured properties for 
models with, for example, very slightly different initial mass \citep{sukhbold_2018_aa}. Our 3D models provide an estimate for the 
variation of the compactness parameter due to different grid resolution and symmetry but should be 
considered in the larger context of 1D stellar models and the other factors that can also contribute to variations 
in key properties.}

\subsection{Comparison between the 1- and 3D Simulations}
In this subsection we  compare the angle-averaged properties of the 
of the 1D \MESA model and the \added{four} 3D models at a time near iron core collapse.
In Figure~\ref{fig:conv_props_1_3d_comparison} we show the convective velocity (top) and 
Mach number (bottom) for these models at $t\approx 420$ s. Considering first the O-shell region, 
situated at a mass coordinate between $m\approx 1.7-2.3$ \msun, the convective velocity speeds
of the \added{four} 3D models agree quite well in shape and magnitude. This is with exception to the 
\texttt{3DOct16km} model for which the velocities in this region are $\approx$ 5-10 km s$^{-1}$ slower. 
In this region, the 1D \MESA model matches the shape of the convective velocity profile somewhat
well but predicts a region with considerable convective activity that is smaller in extent, 
ranging only from $m\approx 1.7-2.0$ \msun.
Additionally, the magnitude of the speeds in this region according to MLT are significantly less,
$\approx 4-5$ times less, than the values found in our 3D models. The Si-shell region is
situated at a mass coordinate of $m\approx 1.55-1.7$ \msun. The convective
 speeds in the \added{three} 3D 32 km models agree well within this region while the \texttt{3DOct16km} model shows a lower
speed of $v_{\rm{conv.}}\approx 160$ km s$^{-1}$ owing to the merging of the convectively burning and
non-convectively burning regions discussed in Section~\ref{sec:3d}. The 1D \MESA model agrees with \texttt{3DOct16km} remarkably well in the shape
and magnitude of the convective velocity speeds in this region with only slight
differences at the outer edge of the Si-shell region being steeper in the 3D model. 
These trends follow a similar behavior when looking at the Mach number profiles. The \MESA model
agrees well in shape and magnitude with the \texttt{3DOct16km} model but significantly underestimates
the values in the O-shell region. 

We consider the core properties in Figure~\ref{fig:core_props_1_3d_comparison} where we show the
specific entropy (top) and electron fraction (bottom) for the same models considered in Figure~\ref{fig:conv_props_1_3d_comparison}. Note that owing to the ``inner boundary condition'' used
for the core, the specific electron fraction for all the models up to a mass coordinate of $m\approx0.8$ 
\msun should be the same value. All models follow a similar specific entropy profile with only minor differences
in mass of the iron core, the mass coordinate where $s \approx 4 \ k_{\rm{B}}$ baryon$^{-1}$. 
Qualitatively, the specific entropy and electron fraction profiles for the 3D simulations are smoother than those of the \MESA model.

\section{Summary and Discussion}
\label{sec:discussion}
We have investigated the long term, multidimensional, hydrodynamical evolution of a 15 \msun star 
for the final seven minutes of Si- and O- shell burning prior to and up to the point of iron core collapse. 
Using the \FLASH simulation framework we evolved eight stellar models at varying resolution, dimensions, and
symmetries to characterize the nature of the convective properties of the stellar models and their implications for 
CCSN explosions.

We find that in general \added{many of} the angle averaged properties of the multidimensional models
\added{were consistent} with predictions made by \MESA. The largest differences observed were found when comparing the convective
velocities in the O-shell region to those in the \MESA model. \added{In this region, the 3D models shows convective 
velocity speeds that were 4-5 times larger and maximum Mach numbers that were approximately a factor of three \added{more} 
than as predicted by \MESA }. \added{This large difference can have quantitative implications for the CCSN mechanism. 
Specifically, the larger velocities which can reduce the needed effective heating behind the stalled shock by instead increasing 
the total stress behind the shock \citep{murphy:2013, couch_2013_aa, couch_2015_ab,muller_2015_ab}. 
For instance, \citet{couch_2015_ab} find that a perturbed 3D progenitor model reduced the need for increased neutrino heating to achieve successful explosion.}
Our 2D models showed a convectively active 
Si-shell region with peak velocities of approximately 500 km s$^{-1}$ near collapse and Mach numbers of $\approx$ 0.1 
near collapse. Within the O-shell region the 2D models show slightly slower convective speeds of $\approx$ 400 km s$^{-1}$ 
and Mach numbers of 0.8-0.12 depending on the resolution of the simulation. The 3D models show velocities 
and Mach numbers lower than this in all
cases. The 4$\pi$ 3D models had convective velocities of $\approx$ 240-260 km s$^{-1}$ in the Si- and O- shell 
moments prior to collapse with Mach values of 0.06. 
\added{However, in our highest resolution 3D model (\texttt{3DOct16km}), the Si-shell region reached a 
lower convective Mach number of $\approx$ 0.03. Our 3D models suggest Si-shell peak Mach numbers 
of 0.03-0.06 while the O-shell values are consistent across all models to 
$\approx$ 5-10\% at $\left< \mathcal{M} \right> \approx 0.06$.}

To characterize the behavior of the convection of the 4$\pi$ 3D \added{models}, we computed power spectra 
of the Si- and O-shell regions by \added{decomposing the velocity field into spherical harmonics} at different times 
throughout the simulation. \added{The spectra computed for the O-shell region showed a peak at $\ell=4$ with 
non-negligible power at even larger scales near collapse. In this region, comparable power was found at intermediate
to small scales of $\ell\approx10-30$ despite the bulk of power being stored at large scales of $\ell \approx 2-4$ near collapse.
The Si-shell region showed a rather broad spectrum at early times with a peak at $\ell\approx30-40$. In the final
$\approx$ seconds of the simulation, we found the spectrum increased significantly in power with the 
peak of the spectrum remaining at $\ell\approx30-40$ but showing an increase in power at $\ell\approx5$.}  

\added{We also consider the impact of grid resolution on our results. We find that our highest resolution 
2D (8 km and 16 km) and $\texttt{3DOct16km}$ models showed different behavior in the Si-shell than the other, less 
resolved models.
This behavior is attributed to a less severe initial and reflected transient in these higher resolution models allowing them 
to maintain better hydrostatic equilibrium (undergo less shell expansion). The subsequent result was a Si-shell 
region that reached higher convective velocities at earlier times than that of their lower resolution 3D counterparts 
leading to the merging of convective and non-convective regions in the Si-shell.
The end result was a wider Si-shell region that gave Mach numbers a factor of two smaller than in the higher resolution 
models. 
The results were found to be independent of the dimensionality as they occurred in the 16 km 3D octant and 2D high 
resolution models.}
However, despite the differences between these evolutionary paths in the Si-shell, the results of the O-shell 
region appear largely unaffected by resolution or geometry, resulting in quantitatively similar properties 
near collapse in all of the 3D models. 
When comparing the \added{four} 3D models for different resolutions and symmetries we also found that the Mach number
profiles in the O-shell region agreed across all models with only a slight difference shown in the 4$\pi$ \added{unperturbed model} 
where larger Mach numbers ($\approx 8 \%$) are found at the base of the O-shell. The Si-shell region Mach number profiles showed 
that the $\texttt{3DOct16km}$ model has a smaller value of $\approx0.03$ while the $\texttt{3DOct32km}$
reached a value approximately twice of that. This difference is again linked to the merging of the convective 
and non-convective regions in the 16 km models.  Another important diagnostic linking the presupernova structure 
to the dynamics of the CCSN mechanism is the compactness parameter. When comparing values of this parameter,
\added{we find that the $4\pi$ 3D models agreed with the value computed for the 1D \MESA model at collapse to 
within $\sim$4\%. However, we show that differences in resolution for the 3D models can lead to a 
$\sim$25\% variation while the assumed octant symmetry could lead to a $\sim$10\% variation. 
The magnitude of these variations suggest that they can be on order of the variations found due to varying 
spatial resolution \citep{sukhbold_2018_aa}, 
size of nuclear reaction network \citep{farmer_2016_aa}, and 
nuclear reaction rate uncertainties of 1D pre-supernova models \citep{fields_2018_aa}.}

In C15, they investigated the final three minutes of Si-shell burning in a 15\msun star evolved 
assuming octant symmetry and a reaction network that included enhanced electron capture rates.
They found that convective speeds in the silicon shell reach values of 80-140 km s$^{-1}$ near collapse. 
These values are approximately a factor of \added{two-three} smaller than what we find in all of our 3D simulations 
and a factor of four smaller than the results suggested by our 2D models. A major cause of these differences
can likely be attributed to the length of their simulation. In all our multidimensional models considered 
here the convective velocity speeds did not increase considerably until about five minutes 
into the simulation. Measuring the turbulent 
kinetic energy power spectrum for their model they found the bulk of the energy residing at small
$l$ values (large scales), at $l\approx4$ due to the imposed octant symmetry. They also found significant
power at an $l\approx10$ value for the Si-shell region near collapse. \added{This is a larger scale
than observed in our 4$\pi$ 3D perturbed model where we observe a peak in the Si-shell 
spectrum at $\ell=30-40$ in the final 20 seconds prior to collapse. }

\citet{muller_2016_aa} investigated the last minutes of O-shell burning in a 18 \msun star. They evolved
the model for $\approx$ five minutes using a contracting inner boundary condition situated at the
base of the O-shell mapped to follow the mass trajectory from the initial \texttt{Kepler} model. In their
simulation of O-shell burning they find transverse velocity speeds that reach values of $\approx$ 250 
km s$^{-1}$ approximately a minute prior to collapse. These values are slightly larger by about 50-100 km s$^{-1}$ than the 
values we find in all of our 3D models at a similar epoch. At the onset of collapse, they observe 
peak Mach numbers in the O-shell of $\left < \mathcal{M} \right > \approx 0.1$ where we 
find a value of $\sim$0.06. They compute the power spectrum for the radial velocity component
into spherical harmonics to characterize the scale of the convection. At the early times, $t=90.9$ s they 
find a similar characteristic scale at $l\approx4$ where the bulk of the power resides.
As the simulation evolves the bulk of the power in their model
shifts to larger scales at $l\approx2-4$.
\added{Near the onset of collapse \citet{muller_2016_aa} observe the emergence of a large scale
mode at $l\approx2$. In our simulation, we see a significant increase in power at $\ell=2$ and $\ell=3$ 
but our peak in the distribution resides at $\ell=4$.}

Recently, \citet{yadav_2019_aa} presented a 4$\pi$ 3D simulation of O-/Ne-shell burning using 
a similar method as presented in \citet{muller_2016_aa}. The simulation was evolved for 420 s 
and captured the merging of a large scale O-Ne shell merger leading to significant deviations
from the properties predicted by the 1D initial model. In this work, they found at $t\approx250$ s,
the barrier separating the O- and Ne-shells disappears due to an increase in entropy in the O-shell 
leading to the merging of the two convective regions. The merger leads to large scale density 
fluctuations characterized by $l\approx1-2$ modes within the merged shell. After the merger
they observe velocity fluctuations on the order of 800 km s$^{-1}$ that increase to as large at 
1600 km s$^{-1}$ near collapse. At collapse they observe Mach numbers of $\approx0.13$ in the 
O/Ne mixed region. These values both suggest that the merger can lead to significantly 
larger deviations from spherical symmetry than as suggested by the model presented in this
work and other simulations of quasi-steady state convection prior to core collapse. 
Despite the merging of the two unique convective regions in the Si-shell observed in most
of our models, we do not observe merging of different burning shell regions in any of our models.

High resolution, long term, 4$\pi$ 3D simulations of CCSN progenitors can provide accurate 
initial conditions for simulations of CCSNe. An accurate representation of the
state of the progenitor prior to collapse can have a favorable impact on the delayed neutrino-driven
explosion mechanism and has important implications for the predictions of key observables from CCSN simulations. 
In addition to fully 3D convection motions, most massive stars are also rotating 
differentially in their cores. 
In the presence of weak seed magnetic fields, this rotation can facilitate a large scale dynamo that 
can have an impact on the progenitor 
evolution and the explosion mechanism. As such, a 
next step in increasing the physics fidelity of supernova progenitor models would be to consider the 
impact of a rotating and magnetic progenitor on the 
observed scale and magnitude of perturbations within the late time burning shell regions. The
direct link between multidimensional rotating and magnetic CCSN progenitors  
and the CCSN mechanism is an important question and is the direction of future work. 

\software{
\MESA \citep[][\url{http://mesa.sourceforge.net}]{paxton_2011_aa,paxton_2013_aa,paxton_2015_aa,paxton_2018_aa},
\MESA \citep[][\url{http://flash.uchicago.edu/site/}]{fryxell_2000_aa},
\texttt{yt} \citep[][\url{https://yt-project.org}]{turk_2011_aa}, and
\texttt{matplotlib} \citep[][\url{https://matplotlib.org}]{hunter_2007_aa},
\texttt{SHTns} \citep{sht_2013_aa} .}

\acknowledgements
We thank 
S. Jones,
P. Grete,
J. Ranta,
P. N. Sagan,
and
M. Zingale 
for useful discussions. 
We thank the anonymous referee for helping significantly improve a previous version of this manuscript.
C.E.F. acknowledges support from a Predoctoral Fellowship administered by the 
National Academies of Sciences, Engineering, and Medicine on behalf of the 
Ford Foundation, an Edward J Petry Graduate Fellowship from Michigan State 
University, and the National Science Foundation Graduate Research Fellowship 
Program under grant number DGE1424871.
SMC is supported by the U.S. Department of Energy, Office of Science, Office of Nuclear Physics, 
under Award Numbers DE-SC0015904 and DE-SC0017955. 
This work was supported in part by Michigan State University through 
computational resources provided by the Institute for Cyber-Enabled Research.
This research made extensive use of the SAO/NASA Astrophysics Data System (ADS).

\bibliographystyle{aasjournal}
\bibliography{prog3dn}

\begin{thebibliography}{}
\expandafter\ifx\csname natexlab\endcsname\relax\def\natexlab#1{#1}\fi
\providecommand{\url}[1]{\href{#1}{#1}}
\providecommand{\dodoi}[1]{doi:~\href{http://doi.org/#1}{\nolinkurl{#1}}}
\providecommand{\doeprint}[1]{\href{http://ascl.net/#1}{\nolinkurl{http://ascl.net/#1}}}
\providecommand{\doarXiv}[1]{\href{https://arxiv.org/abs/#1}{\nolinkurl{https://arxiv.org/abs/#1}}}

\bibitem[{{Arnett}(1994)}]{arnett_1994_aa}
{Arnett}, D. 1994, \apj, 427, 932, \dodoi{10.1086/174199}

\bibitem[{{Arnett} {et~al.}(2009){Arnett}, {Meakin}, \&
  {Young}}]{arnett_2009_aa}
{Arnett}, D., {Meakin}, C., \& {Young}, P.~A. 2009, \apj, 690, 1715.
\newblock \doarXiv{0809.1625}

\bibitem[{{Arnett} \& {Meakin}(2011)}]{arnett_2011_ab}
{Arnett}, W.~D., \& {Meakin}, C. 2011, \apj, 733, 78,
  \dodoi{10.1088/0004-637X/733/2/78}

\bibitem[{{B{\"o}hm-Vitense}(1958)}]{bohm_1958_aa}
{B{\"o}hm-Vitense}, E. 1958, \zap, 46, 108

\bibitem[{{Botticella} {et~al.}(2012){Botticella}, {Smartt}, {Kennicutt},
  {Cappellaro}, {Sereno}, \& {Lee}}]{botticella_2012_aa}
{Botticella}, M.~T., {Smartt}, S.~J., {Kennicutt}, R.~C., {et~al.} 2012, \aap,
  537, A132, \dodoi{10.1051/0004-6361/201117343}

\bibitem[{{C{\^o}t{\'e}} {et~al.}(2017){C{\^o}t{\'e}}, {O'Shea}, {Ritter},
  {Herwig}, \& {Venn}}]{cote_2017_aa}
{C{\^o}t{\'e}}, B., {O'Shea}, B.~W., {Ritter}, C., {Herwig}, F., \& {Venn},
  K.~A. 2017, \apj, 835, 128, \dodoi{10.3847/1538-4357/835/2/128}

\bibitem[{{Couch} {et~al.}(2015){Couch}, {Chatzopoulos}, {Arnett}, \&
  {Timmes}}]{couch_2015_aa}
{Couch}, S.~M., {Chatzopoulos}, E., {Arnett}, W.~D., \& {Timmes}, F.~X. 2015,
  \apjl, 808, L21, \dodoi{10.1088/2041-8205/808/1/L21}

\bibitem[{{Couch} {et~al.}(2013){Couch}, {Graziani}, \&
  {Flocke}}]{couch_2013_ab}
{Couch}, S.~M., {Graziani}, C., \& {Flocke}, N. 2013, \apj, 778, 181,
  \dodoi{10.1088/0004-637X/778/2/181}

\bibitem[{{Couch} \& {O'Connor}(2014)}]{couch_2014_aa}
{Couch}, S.~M., \& {O'Connor}, E.~P. 2014, \apj, 785, 123,
  \dodoi{10.1088/0004-637X/785/2/123}

\bibitem[{{Couch} \& {Ott}(2013)}]{couch_2013_aa}
{Couch}, S.~M., \& {Ott}, C.~D. 2013, \apjl, 778, L7,
  \dodoi{10.1088/2041-8205/778/1/L7}

\bibitem[{{Couch} \& {Ott}(2015)}]{couch_2015_ab}
---. 2015, \apj, 799, 5, \dodoi{10.1088/0004-637X/799/1/5}

\bibitem[{{Couch} {et~al.}(2019){Couch}, {Warren}, \&
  {O'Connor}}]{couch_2019_aa}
{Couch}, S.~M., {Warren}, M.~L., \& {O'Connor}, E.~P. 2019, arXiv e-prints,
  arXiv:1902.01340.
\newblock \doarXiv{1902.01340}

\bibitem[{{Cox} \& {Giuli}(1968)}]{cox_1968_aa}
{Cox}, J.~P., \& {Giuli}, R.~T. 1968, {Principles of Stellar Structure} (New
  York: {Gordon \& Breach})

\bibitem[{Dubey {et~al.}(2009)Dubey, Antypas, Ganapathy, Reid, Riley, Sheeler,
  Siegel, \& Weide}]{dubey_2009_aa}
Dubey, A., Antypas, K., Ganapathy, M.~K., {et~al.} 2009, Parallel Computing,
  35, 512 , \dodoi{https://doi.org/10.1016/j.parco.2009.08.001}

\bibitem[{{Farmer} {et~al.}(2016){Farmer}, {Fields}, {Petermann}, {Dessart},
  {Cantiello}, {Paxton}, \& {Timmes}}]{farmer_2016_aa}
{Farmer}, R., {Fields}, C.~E., {Petermann}, I., {et~al.} 2016, \apjs, 227, 22,
  \dodoi{10.3847/1538-4365/227/2/22}

\bibitem[{{Farmer} {et~al.}(2015){Farmer}, {Fields}, \&
  {Timmes}}]{farmer_2015_aa}
{Farmer}, R., {Fields}, C.~E., \& {Timmes}, F.~X. 2015, \apj, 807, 184,
  \dodoi{10.1088/0004-637X/807/2/184}

\bibitem[{{Fields} {et~al.}(2018){Fields}, {Timmes}, {Farmer}, {Petermann},
  {Wolf}, \& {Couch}}]{fields_2018_aa}
{Fields}, C.~E., {Timmes}, F.~X., {Farmer}, R., {et~al.} 2018, \apjs, 234, 19,
  \dodoi{10.3847/1538-4365/aaa29b}

\bibitem[{{Fryxell} {et~al.}(2000){Fryxell}, {Olson}, {Ricker}, {Timmes},
  {Zingale}, {Lamb}, {MacNeice}, {Rosner}, {Truran}, \&
  {Tufo}}]{fryxell_2000_aa}
{Fryxell}, B., {Olson}, K., {Ricker}, P., {et~al.} 2000, \apjs, 131, 273,
  \dodoi{10.1086/317361}

\bibitem[{{Glas} {et~al.}(2019){Glas}, {Just}, {Janka}, \&
  {Obergaulinger}}]{glas:2019}
{Glas}, R., {Just}, O., {Janka}, H.~T., \& {Obergaulinger}, M. 2019, \apj, 873,
  45

\bibitem[{{Hanke} {et~al.}(2013){Hanke}, {M{\"u}ller}, {Wongwathanarat},
  {Marek}, \& {Janka}}]{hanke:2013}
{Hanke}, F., {M{\"u}ller}, B., {Wongwathanarat}, A., {Marek}, A., \& {Janka},
  H.-T. 2013, \apj, 770, 66

\bibitem[{{Heger} {et~al.}(2000){Heger}, {Langer}, \&
  {Woosley}}]{heger_2000_aa}
{Heger}, A., {Langer}, N., \& {Woosley}, S.~E. 2000, \apj, 528, 368

\bibitem[{{Heger} \& {Woosley}(2010)}]{heger_2010_aa}
{Heger}, A., \& {Woosley}, S.~E. 2010, \apj, 724, 341,
  \dodoi{10.1088/0004-637X/724/1/341}

\bibitem[{{Hopkins} {et~al.}(2011){Hopkins}, {Quataert}, \&
  {Murray}}]{hopkins_2011_aa}
{Hopkins}, P.~F., {Quataert}, E., \& {Murray}, N. 2011, \mnras, 417, 950,
  \dodoi{10.1111/j.1365-2966.2011.19306.x}

\bibitem[{Hunter(2007)}]{hunter_2007_aa}
Hunter, J.~D. 2007, Computing In Science \&amp; Engineering, 9, 90

\bibitem[{{Janka}(2012)}]{janka_2012_aa}
{Janka}, H.-T. 2012, Annual Review of Nuclear and Particle Science, 62, 407,
  \dodoi{10.1146/annurev-nucl-102711-094901}

\bibitem[{{Jones} {et~al.}(2016){Jones}, {Andrassy}, {Sandalski}, {Davis},
  {Woodward}, \& {Herwig}}]{jones_2016_aa}
{Jones}, S., {Andrassy}, R., {Sandalski}, S., {et~al.} 2016, ArXiv e-prints.
\newblock \doarXiv{1605.03766}

\bibitem[{{Kraichnan}(1967)}]{kraichnan:1967}
{Kraichnan}, R.~H. 1967, Physics of Fluids, 10, 1417

\bibitem[{{Lai} \& {Goldreich}(2000)}]{lai:2000}
{Lai}, D., \& {Goldreich}, P. 2000, \apj, 535, 402

\bibitem[{{Langanke} \& {Mart{\'{\i}}nez-Pinedo}(2000)}]{langanke_2000_aa}
{Langanke}, K., \& {Mart{\'{\i}}nez-Pinedo}, G. 2000, Nuclear Physics A, 673,
  481, \dodoi{10.1016/S0375-9474(00)00131-7}

\bibitem[{Lee \& Deane(2009)}]{lee_2008_aa}
Lee, D., \& Deane, A.~E. 2009, Journal of Computational Physics, 228, 952 ,
  \dodoi{https://doi.org/10.1016/j.jcp.2008.08.026}

\bibitem[{{Lentz} {et~al.}(2015){Lentz}, {Bruenn}, {Hix}, {Mezzacappa},
  {Messer}, {Endeve}, {Blondin}, {Harris}, {Marronetti}, \&
  {Yakunin}}]{lentz:2015}
{Lentz}, E.~J., {Bruenn}, S.~W., {Hix}, W.~R., {et~al.} 2015, \apjl, 807, L31

\bibitem[{{Mabanta} \& {Murphy}(2018)}]{mabanta_2018_aa}
{Mabanta}, Q.~A., \& {Murphy}, J.~W. 2018, \apj, 856, 22,
  \dodoi{10.3847/1538-4357/aaaec7}

\bibitem[{{Mazurek} {et~al.}(1974){Mazurek}, {Truran}, \&
  {Cameron}}]{mazurek_1974_aa}
{Mazurek}, T.~J., {Truran}, J.~W., \& {Cameron}, A.~G.~W. 1974, \apss, 27, 261,
  \dodoi{10.1007/BF00643877}

\bibitem[{{Meakin} \& {Arnett}(2007)}]{meakin_2007_aa}
{Meakin}, C.~A., \& {Arnett}, D. 2007, \apj, 665, 690, \dodoi{10.1086/519372}

\bibitem[{Muller \& Janka(2015)}]{muller_2015_ab}
Muller, B., \& Janka, H.-T. 2015, Monthly Notices of the Royal Astronomical
  Society, 448, 2141, \dodoi{10.1093/mnras/stv101}

\bibitem[{{M{\"u}ller} {et~al.}(2017){M{\"u}ller}, {Melson}, {Heger}, \&
  {Janka}}]{muller:2017}
{M{\"u}ller}, B., {Melson}, T., {Heger}, A., \& {Janka}, H.-T. 2017, \mnras,
  472, 491

\bibitem[{{M{\"u}ller} {et~al.}(2016){M{\"u}ller}, {Viallet}, {Heger}, \&
  {Janka}}]{muller_2016_aa}
{M{\"u}ller}, B., {Viallet}, M., {Heger}, A., \& {Janka}, H.-T. 2016, ArXiv
  e-prints.
\newblock \doarXiv{1605.01393}

\bibitem[{{Murphy} {et~al.}(2013){Murphy}, {Dolence}, \&
  {Burrows}}]{murphy:2013}
{Murphy}, J.~W., {Dolence}, J.~C., \& {Burrows}, A. 2013, \apj, 771, 52

\bibitem[{{Nagakura} {et~al.}(2019){Nagakura}, {Burrows}, {Radice}, \&
  {Vartanyan}}]{nagakura_2019_aa}
{Nagakura}, H., {Burrows}, A., {Radice}, D., \& {Vartanyan}, D. 2019, arXiv
  e-prints.
\newblock \doarXiv{1905.03786}

\bibitem[{{O'Connor} \& {Ott}(2011)}]{oconnor_2011_aa}
{O'Connor}, E., \& {Ott}, C.~D. 2011, \apj, 730, 70,
  \dodoi{10.1088/0004-637X/730/2/70}

\bibitem[{{O'Connor} \& {Couch}(2018{\natexlab{a}})}]{oconnor_2018_ab}
{O'Connor}, E.~P., \& {Couch}, S.~M. 2018{\natexlab{a}}, \apj, 865, 81,
  \dodoi{10.3847/1538-4357/aadcf7}

\bibitem[{{O'Connor} \& {Couch}(2018{\natexlab{b}})}]{oconnor_2018_aa}
---. 2018{\natexlab{b}}, \apj, 854, 63, \dodoi{10.3847/1538-4357/aaa893}

\bibitem[{{{\"O}zel} {et~al.}(2012){{\"O}zel}, {Psaltis}, {Narayan}, \& {Santos
  Villarreal}}]{ozel_2012_aa}
{{\"O}zel}, F., {Psaltis}, D., {Narayan}, R., \& {Santos Villarreal}, A. 2012,
  \apj, 757, 55, \dodoi{10.1088/0004-637X/757/1/55}

\bibitem[{{Paxton} {et~al.}(2011){Paxton}, {Bildsten}, {Dotter}, {Herwig},
  {Lesaffre}, \& {Timmes}}]{paxton_2011_aa}
{Paxton}, B., {Bildsten}, L., {Dotter}, A., {et~al.} 2011, \apjs, 192, 3,
  \dodoi{10.1088/0067-0049/192/1/3}

\bibitem[{{Paxton} {et~al.}(2013){Paxton}, {Cantiello}, {Arras}, {Bildsten},
  {Brown}, {Dotter}, {Mankovich}, {Montgomery}, {Stello}, {Timmes}, \&
  {Townsend}}]{paxton_2013_aa}
{Paxton}, B., {Cantiello}, M., {Arras}, P., {et~al.} 2013, \apjs, 208, 4,
  \dodoi{10.1088/0067-0049/208/1/4}

\bibitem[{{Paxton} {et~al.}(2015){Paxton}, {Marchant}, {Schwab}, {Bauer},
  {Bildsten}, {Cantiello}, {Dessart}, {Farmer}, {Hu}, {Langer}, {Townsend},
  {Townsley}, \& {Timmes}}]{paxton_2015_aa}
{Paxton}, B., {Marchant}, P., {Schwab}, J., {et~al.} 2015, \apjs, 220, 15,
  \dodoi{10.1088/0067-0049/220/1/15}

\bibitem[{{Paxton} {et~al.}(2018){Paxton}, {Schwab}, {Bauer}, {Bildsten},
  {Blinnikov}, {Duffell}, {Farmer}, {Goldberg}, {Marchant}, {Sorokina},
  {Thoul}, {Townsend}, \& {Timmes}}]{paxton_2018_aa}
{Paxton}, B., {Schwab}, J., {Bauer}, E.~B., {et~al.} 2018, \apjs, 234, 34,
  \dodoi{10.3847/1538-4365/aaa5a8}

\bibitem[{{Paxton} {et~al.}(2019){Paxton}, {Smolec}, {Gautschy}, {Bildsten},
  {Cantiello}, {Dotter}, {Farmer}, {Goldberg}, {Jermyn}, {Kanbur}, {Marchant},
  {Schwab}, {Thoul}, {Townsend}, {Wolf}, {Zhang}, \& {Timmes}}]{paxton_2019_aa}
{Paxton}, B., {Smolec}, R., {Gautschy}, A., {et~al.} 2019, arXiv e-prints.
\newblock \doarXiv{1903.01426}

\bibitem[{{Pignatari} {et~al.}(2016){Pignatari}, {Herwig}, {Hirschi},
  {Bennett}, {Rockefeller}, {Fryer}, {Timmes}, {Ritter}, {Heger}, {Jones},
  {Battino}, {Dotter}, {Trappitsch}, {Diehl}, {Frischknecht}, {Hungerford},
  {Magkotsios}, {Travaglio}, \& {Young}}]{pignatari_2016_aa}
{Pignatari}, M., {Herwig}, F., {Hirschi}, R., {et~al.} 2016, \apjs, 225, 24,
  \dodoi{10.3847/0067-0049/225/2/24}

\bibitem[{{Radice} {et~al.}(2015){Radice}, {Couch}, \& {Ott}}]{radice:2015}
{Radice}, D., {Couch}, S.~M., \& {Ott}, C.~D. 2015, Computational Astrophysics
  and Cosmology, 2, 7

\bibitem[{{Radice} {et~al.}(2016){Radice}, {Ott}, {Abdikamalov}, {Couch},
  {Haas}, \& {Schnetter}}]{radice:2016}
{Radice}, D., {Ott}, C.~D., {Abdikamalov}, E., {et~al.} 2016, \apj, 820, 76

\bibitem[{{Rembiasz} {et~al.}(2017){Rembiasz}, {Obergaulinger},
  {Cerd{\'a}-Dur{\'a}n}, {Aloy}, \& {M{\"u}ller}}]{rembiasz:2017}
{Rembiasz}, T., {Obergaulinger}, M., {Cerd{\'a}-Dur{\'a}n}, P., {Aloy},
  M.-{\'A}., \& {M{\"u}ller}, E. 2017, \apjs, 230, 18

\bibitem[{{Roberts} {et~al.}(2016){Roberts}, {Ott}, {Haas}, {O'Connor},
  {Diener}, \& {Schnetter}}]{roberts_2016_aa}
{Roberts}, L.~F., {Ott}, C.~D., {Haas}, R., {et~al.} 2016, \apj, 831, 98,
  \dodoi{10.3847/0004-637X/831/1/98}

\bibitem[{Schaeffer(2013)}]{sht_2013_aa}
Schaeffer, N. 2013, Geochemistry, Geophysics, Geosystems, 14, 751,
  \dodoi{10.1002/ggge.20071}

\bibitem[{{Su} {et~al.}(2018){Su}, {Hopkins}, {Hayward}, {Ma},
  {Boylan-Kolchin}, {Kasen}, {Kere{\v{s}}}, {Faucher-Gigu{\`e}re}, {Orr}, \&
  {Wheeler}}]{su_2018_aa}
{Su}, K.-Y., {Hopkins}, P.~F., {Hayward}, C.~C., {et~al.} 2018, \mnras, 480,
  1666, \dodoi{10.1093/mnras/sty1928}

\bibitem[{{Sukhbold} {et~al.}(2016){Sukhbold}, {Ertl}, {Woosley}, {Brown}, \&
  {Janka}}]{sukhbold_2016_aa}
{Sukhbold}, T., {Ertl}, T., {Woosley}, S.~E., {Brown}, J.~M., \& {Janka}, H.-T.
  2016, \apj, 821, 38, \dodoi{10.3847/0004-637X/821/1/38}

\bibitem[{{Sukhbold} \& {Woosley}(2014)}]{sukhbold_2014_aa}
{Sukhbold}, T., \& {Woosley}, S.~E. 2014, \apj, 783, 10,
  \dodoi{10.1088/0004-637X/783/1/10}

\bibitem[{{Sukhbold} {et~al.}(2018){Sukhbold}, {Woosley}, \&
  {Heger}}]{sukhbold_2018_aa}
{Sukhbold}, T., {Woosley}, S.~E., \& {Heger}, A. 2018, \apj, 860, 93,
  \dodoi{10.3847/1538-4357/aac2da}

\bibitem[{{Timmes} {et~al.}(2000){Timmes}, {Hoffman}, \&
  {Woosley}}]{timmes_2000_ab}
{Timmes}, F.~X., {Hoffman}, R.~D., \& {Woosley}, S.~E. 2000, \apjs, 129, 377

\bibitem[{{Timmes} \& {Swesty}(2000)}]{timmes_2000_aa}
{Timmes}, F.~X., \& {Swesty}, F.~D. 2000, \apjs, 126, 501

\bibitem[{{Timmes} {et~al.}(1995){Timmes}, {Woosley}, \&
  {Weaver}}]{timmes_1995_aa}
{Timmes}, F.~X., {Woosley}, S.~E., \& {Weaver}, T.~A. 1995, \apjs, 98, 617

\bibitem[{Toro(1999)}]{toro_1999_aa}
Toro, E.~F. 1999, Riemann Solvers and Numerical Methods for Fluid Dynamics
  (Springer, Berlin, Heidelberg)

\bibitem[{{Trampedach} {et~al.}(2014){Trampedach}, {Stein},
  {Christensen-Dalsgaard}, {Nordlund}, \& {Asplund}}]{trampedach_2014_aa}
{Trampedach}, R., {Stein}, R.~F., {Christensen-Dalsgaard}, J., {Nordlund},
  {\AA}., \& {Asplund}, M. 2014, \mnras, 445, 4366,
  \dodoi{10.1093/mnras/stu2084}

\bibitem[{{Turk} {et~al.}(2011){Turk}, {Smith}, {Oishi}, {Skory}, {Skillman},
  {Abel}, \& {Norman}}]{turk_2011_aa}
{Turk}, M.~J., {Smith}, B.~D., {Oishi}, J.~S., {et~al.} 2011, \apjs, 192, 9,
  \dodoi{10.1088/0067-0049/192/1/9}

\bibitem[{{Vartanyan} {et~al.}(2019){Vartanyan}, {Burrows}, {Radice},
  {Skinner}, \& {Dolence}}]{vartanyan:2019}
{Vartanyan}, D., {Burrows}, A., {Radice}, D., {Skinner}, M.~A., \& {Dolence},
  J. 2019, \mnras, 482, 351

\bibitem[{{Viallet} {et~al.}(2013){Viallet}, {Meakin}, {Arnett}, \&
  {Moc{\'a}k}}]{viallet_2013_aa}
{Viallet}, M., {Meakin}, C., {Arnett}, D., \& {Moc{\'a}k}, M. 2013, \apj, 769,
  1, \dodoi{10.1088/0004-637X/769/1/1}

\bibitem[{{Woosley} \& {Heger}(2007)}]{woosley_2007_aa}
{Woosley}, S.~E., \& {Heger}, A. 2007, \physrep, 442, 269,
  \dodoi{10.1016/j.physrep.2007.02.009}

\bibitem[{{Woosley} \& {Heger}(2015)}]{woosley_2015_aa}
---. 2015, \apj, 810, 34, \dodoi{10.1088/0004-637X/810/1/34}

\bibitem[{Woosley {et~al.}(2002)Woosley, Heger, \& Weaver}]{woosley_2002_aa}
Woosley, S.~E., Heger, A., \& Weaver, T.~A. 2002, Rev. Mod. Phys., 74, 1015,
  \dodoi{10.1103/RevModPhys.74.1015}

\bibitem[{{Yadav} {et~al.}(2020){Yadav}, {M{\"u}ller}, {Janka}, {Melson}, \&
  {Heger}}]{yadav_2019_aa}
{Yadav}, N., {M{\"u}ller}, B., {Janka}, H.~T., {Melson}, T., \& {Heger}, A.
  2020, \apj, 890, 94, \dodoi{10.3847/1538-4357/ab66bb}

\bibitem[{{Zingale} {et~al.}(2002){Zingale}, {Dursi}, {ZuHone}, {Calder},
  {Fryxell}, {Plewa}, {Truran}, {Caceres}, {Olson}, {Ricker}, {Riley},
  {Rosner}, {Siegel}, {Timmes}, \& {Vladimirova}}]{zingale_2002_aa}
{Zingale}, M., {Dursi}, L.~J., {ZuHone}, J., {et~al.} 2002, \apjs, 143, 539,
  \dodoi{10.1086/342754}

\end{thebibliography}

\end{document}